\documentclass{mn2e}

\usepackage{psfig,epsfig,amssymb,times}

\newcommand{\rashort}[2]{$#1^{\mathrm{h}}#2^{\mathrm{m}}$}
\newcommand{\arcsd}[2]{$#1^{\prime\prime}\!\!.#2$}           
\newcommand{\arcmn}[2]{$#1^{\prime}\!\!.#2$}             

\newcommand{\etal}{{et al}\/. }
\newcommand{\eg}{{\em eg: }}
\newcommand{\ie}{{\em ie: }}

\newcommand{\msun}{\hbox{M$_{\odot}$}}

\title{The cluster environments of radio-loud quasars at $0.6<z<1.1$}
\author[J. M. Barr et al.]
       {J. M. Barr,$^1$\thanks{E-mail: j.barr@bristol.ac.uk}
	M. N. Bremer,$^{1}$
	J. C. Baker$^{2}$
 	 and M. D. Lehnert$^{3}$ \\
	$^1$Department of Physics, University of Bristol, 
	H.H. Wills Laboratory, Tyndall Avenue, Bristol, BS8 1TL, U.K.\\
	$^2$University of Oxford, Astrophysics Department, Keble Road, 
	Oxford, OX1 3RH, U.K.\\
	$^3$Max-Planck-Institut f\"ur extraterrestriche Physik, 
        Giessenbachstra\ss e, 85748 Garching bei M\"unchen, Germany\\
}

\date{\today}

\pagerange{}
\pubyear{2001}

\begin{document}

\maketitle

\label{firstpage}

\begin{abstract}

We have carried out multicolour imaging of a complete sample of
radio-loud quasars at $0.6<z<1.1$ and find groups or clusters of
galaxies in the fields of at least 8 and possibly 13 of the 21
sources. There is no evidence for an evolution in the richness of the
environments of radio-loud quasars from other low-redshift studies to
$z \gtrsim 0.9$. The quasars associated with groups and clusters in
our sample do not necessarily reside in the centre of the galaxy
distribution which rarely displays a spherical geometry.  Clustering
is preferentially associated with small or asymmetric steep-spectrum
radio sources. The quasars with the largest projected angular size
are, in nearly all cases, found in non-clustered environments.
Radio-based selection (including source size) of high-redshift groups
and clusters can be a very efficient method of detecting rich
environments at these redshifts.

We find that in optical searches for galaxy overdensities above $z
\sim 0.6$ multiple filters must be used. If the single-filter counting
statistics used by groups at lower redshift are applied to our data,
uncertainties are too large to make accurate quantifications of
cluster richness. This means that genuine clustering of galaxies about
quasars will be missed and, in $\sim 10\%$ of cases, putative clusters
turn out to be false detections. The statistics are further diluted by
the fact that galaxy overdensities are generally not centred on the
quasar.

\end{abstract}


\section{Introduction}

Groups and clusters of galaxies at high redshift are key targets for
studying the evolution of galaxies and large-scale structure. They
provide insight into the star formation history of the earliest
galaxies (\eg Barger \etal 1996; Poggianti \etal 1996; Balogh \etal
1997; Kodama \& Bower
2001)\nocite{barger96,poggianti96,balogh97,kodama01}, as well as being
powerful probes of the growth of structure
\cite{dressler80,vandokkum01}.  In addition, they are observational
points of reference for theoretical simulations of structure
formation which can provide insight into the cosmological parameters
(\eg Gross \etal 1998; Jenkins \etal 1998; Yoshida \etal
2001)\nocite{gross98,jenkins98,yoshida01}.  However, finding large
numbers of clustered systems above $z \sim 0.5$ has proved difficult
and time consuming. The main methods focus on searching for
large-scale or extended structure over wide areas. One such technique
has looked for X-ray emission from hot intracluster gas (\eg Gioia \&
Luppino 1994; Rosati \etal 1998\nocite{rosati98,gioia94}), which is
biassed toward finding the most massive and/or merging systems.  Other
efforts have concentrated on finding overdensities of objects in deep
optical images taken in a single filter (\eg Couch \etal 1991; Lidman
\& Petersen 1996; Postman \etal 1996; Olsen \etal 1999{\em
a}\nocite{couch91,lidman96,postman96,olsen99a}). Optical
identification of clusters on the basis of numerical/statistical
overdensities of faint galaxies becomes increasingly problematic at
higher redshift ($z > 0.5$) due to the variance in the faint number
counts on the scale length of clusters. More recent surveys have
attempted to mitigate this latter problem by using multiple filters to
preselect objects with the same colours , and hence probable redshift
(\eg Olsen \etal 1999{\em b}; Gladders \& Yee 2000; Martini 2000;
Drory \etal
2001)\nocite{olsen99b,gladders00,martini01,drory01}. However, all
blank-field surveys are still inefficient in terms of clusters found
per unit observing time.

It has been known for some time that groups and clusters of galaxies
share a close relationship with active galactic nuclei (AGN). AGN at
low redshift have been shown to exist in environments of above average
galactic density \cite{seldner78,longair79,steiner82,yee84,yee87} and
studies extending these observations to higher redshift have shown
that at $z > 0.3$ high-luminosity AGN reside in the richest clusters
\cite{ellingson91,hill91}. Detailed knowledge of an AGN environment
can provide tests of proposed unification schemes \cite{wurtz97}, as
well as shedding light on the luminosity evolution of AGN
\cite{ellingson91}. Furthermore, the current co-moving space density
of Abell class 0 clusters at low redshift is similar to that of
powerful radio sources at $z \sim 2-3$ \cite{mclure01}. The
implication is that all present day clusters hosted a powerful radio
source at sometime in the past (see also Bremer \etal
1992)\nocite{bremer92}. We exploit this close relationship in order to
create a sample of distant groups and clusters of galaxies with well
understood biases by searching for overdensities of galaxies
associated with radio-loud quasars (RLQs).

\begin{figure}

  \begin{center}

    \epsfig{file=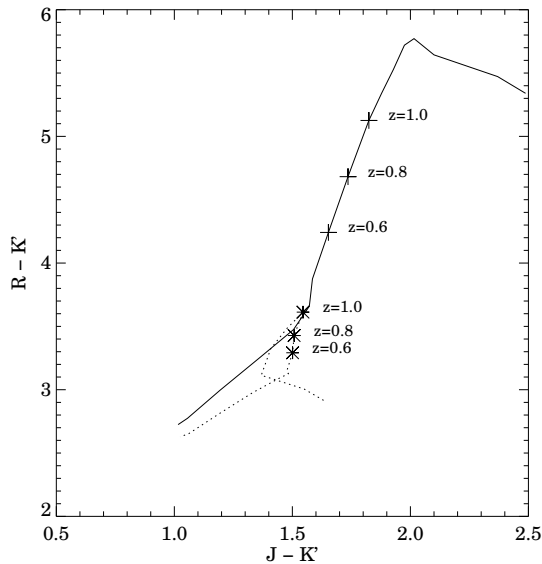,angle=0,width=8cm}

    \caption{The evolution of $R-K^{\prime}$ {\em vs.} $J-K^{\prime}$
    colours for two galaxies at redshifts 0 to 2. Thick line: An
    elliptical formed from a single burst of star formation at
    $z=5$. Dotted line: A spiral formed at the same $z$. Indicated are
    the redshifts at points relevant to our sample. The plots were
    constructed using the synthetic spectra of Fioc \&
    Rocca-Volmerange (1997).}

    \label{fig:coltab}

  \end{center}

\end{figure}

Previous studies of clustering of galaxies around AGN have endeavoured
to quantify the clustering with reference to the statistical
overdensity of objects found within a certain radius (usually 0.5 Mpc)
of the AGN in a single filter (\eg Yee \& Green 1987\nocite{yee87};
Hutchings 1995\nocite{hutchings95}; Hall, Green \& Cohen
1998\nocite{hall98a}; Wold \etal 2000\nocite{wold00}). However, a more
robust strategy, which we follow, is to identify high-redshift cluster
members from their optical and near-infrared (NIR) colours. This
exploits the narrow colour dispersion amongst the passively-evolving
elliptical galaxy population (the red sequence) making up the cores of
clusters \cite{baum59,bower92,stanford98}. A similar method is used in
the wide-field cluster searches of Gladders \& Yee
(2000)\nocite{gladders00} who claim that the red sequence is present
as a visible indicator of a cluster core out to $z > 1.2$. Our method
has the advantage of being able to select the redshift of interest and
hence choose filters to pick out the characteristic cluster galaxies
at that redshift. Core cluster members above $z \sim 0.5$ are readily
identifiable by using a combination of optical and NIR
filters. Optical filters can be used to straddle the prominent 4000\AA
\ break of passively-evolving elliptical galaxies, thus maximising the
contrast for these objects. Using $J$ and $K$ filters with one optical
filter such as $R$ or $I$ separates high-redshift ellipticals from
spirals or M-stars and brown dwarfs, as shown in Pozzetti \& Manucci
(2000)\nocite{pozzetti00} and Figure \ref{fig:coltab}.

\nocite{fioc97}

This paper presents results of deep multicolour imaging of a
representative sample of 21 radio-loud quasars with $0.6 < z <
1.1$. The goal is to identify evolved cluster cores by finding
excesses of passively-evolving elliptical galaxies in the quasar
fields.  Preliminary results for two fields have been published in
Baker \etal (2001)\nocite{baker01} and Bremer, Baker \& Lehnert
(2002)\nocite{bremer02}. Here, as in these previous papers, we analyse
quasar fields individually and report the identification of clustering
in the fields of up to 13 of the 21 RLQs. However, the power of a
complete redshift-limited target sample is that we can extend the
analysis to the sample as a whole. As well as discovering
high-redshift groups and clusters of galaxies, we can correlate the
existence of clustering with the properties of the RLQs. A detailed
analysis of how the radio properties of the sources depend on
environment, incorporating theoretical considerations of the growth of
FRII radio lobes in clustered environments will be published in a
subsequent paper (Barr, Bremer \& Baker 2003, in preparation). In the
present analysis, we concentrate on detecting clustered systems about
AGN and the optical-NIR properties of their core galaxy population.

Details of the target sample and observations are given in Section
\ref{sec:obs}. Single filter statistics are measured and compared with
other studies in Section \ref{sec:sing}. Multicolour analysis of all
the fields are presented in Section \ref{sec:mult}. A discussion of
the relative merits of cluster quantification methods is given in
Section \ref{sec:disc}, along with the cluster characteristics of the
RLQ sample. For comparison with previous studies we adopt a $H_{0}=50$
km s$^{-1}$ Mpc$^{-1}$, $\Omega_{m}=1$, $\Lambda = 0$ cosmology.

\section{Observations and Data reduction}
\label{sec:obs}

\subsection{The Sample}

The quasars used for this study are a redshift-limited subsample of
the Molonglo Quasar Sample (MQS; Kapahi \etal 1998)\nocite{kapahi98}
of low-frequency-selected radio-loud quasars. This is a highly
complete ($>97\%$) sample of 111 RLQs with $S_{408}>0.95$ Jy in the
declination range $-30^{\circ} > \delta > -20^{\circ}$, and Galactic
latitude $|b| > 20^{\circ}$ (excluding the R.A. ranges
\rashort{06}{00} -- \rashort{09}{00} and \rashort{14}{03} --
\rashort{20}{20}).  These objects are on average five times less
powerful and so more numerous at a given redshift than 3CR sources
\cite{laing83}. The MQS therefore probes a more representative subset
of the RLQ population than do 3C quasars.

The sample presented in this paper is well suited to a study of
radio-loud quasar environments. Its low-frequency (408 MHz) radio
selection effectively picks out the steep-spectrum lobe contribution
which results in a sample with few beamed objects, \ie those which may
be of intrinsically lower luminosity but have Doppler boosted emission
due to the orientation of the radio jet to the line of
sight. Excluding sources with highly beamed emission is important as
it gives us a good measure of the intrinsic power of the radio
source. This is vital when attempting to correlate radio and
environmental properties. Selecting on radio luminosity also prevents
the omission of optically obscured quasars. Additionally, the MQS has
the advantage of having been studied in a number of wavelength regimes
(\eg Baker, Hunstead \& Brinkmann 1995; Baker 1997; Ishwara-Chandra
1998; Kapahi \etal 1998; Baker \etal 1999; Ishwara-Chandra \etal 2001;
Baker \etal
2002)\nocite{baker95,baker97,ishwara-chandra98,kapahi98,baker99,ishwara-chandra01,baker02}.

We have observed an RA- and redshift-limited subset of the MQS. Within
the range \rashort{0}{0}$< \mathrm{RA} <$\rashort{14}{0} there are 30
MQS objects which have $0.6<z<1.1$; we have data for 21 of these. The
other nine sources were {\em randomly} excluded due to observing time
restrictions. No quasars were excluded based on optical or radio
criteria. The RLQs in this paper can therefore be thought of as a
representative sample.

\subsection{Optical images}

\begin{table*}
\caption{Summary of optical and NIR observing runs for the sample.}
\begin{tabular}{llllcllc}

\hline\hline

Run & Date & Telescope & Instrument & Pixel Scale & Field of View & Filters & Average Seeing  \\

\hline

A & 1996 Feb 14-16 & ESO 2.2m & IRAC2B & \arcsd{0}{51} & \arcmn{2}{17}
$\times$ \arcmn{2}{17} & $J$,$K^{\prime}$ & \arcsd{0}{8} \\

B & 1996 Mar 19-21 & CTIO 4m & PFCCD & \arcsd{0}{16} & \arcmn{5}{4}
$\times$ \arcmn{5}{4} & $B$,$I$ & \arcsd{0}{5} \\

C & 1997 Mar 12-15 & ESO 3.6m & EFOSC & \arcsd{0}{61} & \arcmn{5}{2}
$\times$ \arcmn{5}{2} & $V$,$R$,$I$,$z$ & \arcsd{0}{7} \\

D & 1997 Mar 17-22 & ESO 2.2m & IRAC2B & \arcsd{0}{51} & \arcmn{2}{17}
$\times$ \arcmn{2}{17} &$J$,$K^{\prime}$ & \arcsd{1}{1} \\

E & 1998 Feb 14-16 & CTIO 4m & CIRIM & \arcsd{0}{22} & \arcmn{0}{9}
$\times$ \arcmn{0}{9} & $H$ & \arcsd{0}{7} \\

F & 1998 Mar 31-Apr 1 & ESO 3.6m & EFOSC2 & \arcsd{0}{30} &
\arcmn{5}{2} $\times$ \arcmn{5}{2} & $R$,$I$ & \arcsd{0}{8} \\

G & 1999 Sep 9 & AAT 3.9m & Taurus & \arcsd{0}{37} & \arcmn{9}{8} diameter &
$V$,$R$,$I$ & \arcsd{1}{5} \\

H & 2000 Apr 11 & AAT 3.9m & Taurus & \arcsd{0}{37} & \arcmn{9}{8}
diameter & $V$,$R$,$I$ & \arcsd{2}{0} \\

\hline
\end{tabular}
\label{tab:runs}
\end{table*}

Optical multicolour imaging of the fields was carried out in four
separate runs using the PFCCD on the CTIO 4m Blanco telescope, EFOSC
and EFOSC2 on the ESO 3.6m at La Silla and Taurus on the AAT. The
field of view was $\sim 5^{\prime}$ for the CTIO and ESO runs and
\arcmn{9}{8} for the AAT data. Pixel scales ranged from \arcsd{0}{16}
for PFCCD to \arcsd{0}{61} on EFOSC. Seeing varied between
\arcsd{0}{5} and \arcsd{2}{0}. The observing setup and conditions for
all runs are detailed in Table~\ref{tab:runs}.

The quasar fields were observed in two or more filters, selected from
$B$, $V$, $R$, $I$ and Gunn $z$ according to the redshift of the
target.  The filters were chosen to straddle the 4000\AA \ break and
thus maximise the contrast for early type galaxies. Observations of
each field in each filter were made up of 2 or 3 integrations dithered
by $\sim 10^{\prime\prime}$ with respect to each other to facilitate
cosmic ray rejection. Total integration times were typically 30
minutes to 1 hour per filter. Landolt photometric standards
\cite{landolt92} were observed on 2--3 occasions on each night. Full
details of the filters and exposure times used are given in
Table~\ref{tab:obs}.

\begin{table*}
\caption{Summary of observations for the sample. For each object
exposure times in seconds are on the upper row and the completeness
limit in magnitudes is on the lower row.}
\begin{tabular}{lclccccclcccc}

\hline\hline

MRC quasar & & \multicolumn{6}{c}{Optical} & & \multicolumn{4}{c}{NIR} \\

\cline{3-8}\cline{10-13}\\

& & Run(s) & $B$ & $V$ & $R$ & $I$ & $z$ & & Run(s) & $J$ & $H$ & $K^{\prime}$ \\ 

\hline

B0030--220 & & G & & 1800 & & 1800 & \\

 & & & & 24.0 & & 22.5 & \\

B0058--229 & & G & & 1800 & & 1800 & \\

 & & & & 23.0 & & 21.5 & \\

B0209--237 & & G & & 1800 & & 1800 & \\

 & & & & 23.5 & & 21.5 & \\

B0346--279 & & C & & & 1800 & & 1800 & & A & 3600 & & 3600 \\

 & & & & & 23.5 & & 22.0 & & & 20.5 & & 19.0 \\

B0450--221 & & C & & 1800 & & 3600 & & & A & 3600 & & 3600 \\

 & & & & 24.5 & & 23.5 & & & & 20.0 & & 18.5 \\

B0941--200 & & H & & 900 & 900 & 900 & & & A,E & 2400 & 3500 & 4020 \\

 & & & & 23.5 & 22.5 & 21.5 & &  & & 20.0 & 21.0 & 17.5 \\

B1006--299 & & B,C & 1800 & & 1800 & 1800 & 6300 & & A & 2400 & & 3600 \\

 & & & 24.0 & & 24.0 & 22.5 & 22.5 & & & 19.5 & & 18.5 \\
 
B1055--242 & & C & & & 3000 & & 3600 & & D & 4800 & \\

 & & & & & 24.0 & & 22.5 & & & 20.5 \\

B1121--238 & & C & & 900 & 600 & 900 & & & D & 1600 & & 1800 \\

 & & & & 24.5 & 24.0 & 22.5 & & & & 20.0 & & 18.5 \\

B1202--262 & & F & & & 1800 & 1800 & \\

 & & & & & 25.0 & 23.5 \\

B1208--277 & & C & & 1800 & & 1800 & & & A & 3720 & & 3000 \\

 & & & & 24.5 & & 23.0 & & & & 20.0 & & 18.5 \\

B1217--209 & & F & & & 1800 & 1200 & \\

 & & & & & 24.5 & 24.0 \\

B1222--293 & & F & & & 1800 & 1800 & \\

 & & & & & 25.0 & 24.0 \\

B1224--262 & & C & & 1800 & 900 & 1800 & & & D & 3200 & & 1800 \\

 & & & & 24.5 & 24.0 & 23.0 & & & & 20.0 & & 18.5 \\

B1226--297 & & F & & & 1800 & 1800 & \\

 & & & & & 24.5 & 23.5 \\

B1247--290 & & C & & 900 & 600 & 1800 & & & D & & & 2400 \\

 & & & & 24.5 & 24.0 & 23.0 & & & & & & 17.5 \\

B1301--251 & & B,C & 1800 & 1800 & & 900 & 1800 & & E & & 3600 \\

 & & & 24.5 & 24.5 & & 22.0 & 22.5 & & & & 22.0 \\ 

B1303--250 & & F & & & 1200 & 600 & \\

 & & & & & 24.5 & 23.0 \\

B1349--265 & & B,C & 1800 & 1800 & & 1800 & & & D,E & 11200 & 3150 \\

 & & & 24.0 & 24.5 & & 22.0 & & & & 21.5 & 21.0 \\

B1355--236 & & B,C & 1800 & & 600 & 1200 & & & &  \\

 & & & 24.0 & & 24.0 & 22.0 \\

B1359--281 & & B & 1800 & & & 1200 & & & & \\

 & & & 24.0 & & & 22.0 \\

\hline
\end{tabular}
\label{tab:obs}
\end{table*}

The data were bias subtracted and flat-fielded in a standard fashion
using programs written in IDL\footnote{Interactive data language; {\tt
http://www.rsinc.com/idl/index.asp}}. The flat-fielding was
accomplished by using twilight sky flats or median-summed images of
the science frames. Images in each filter were registered and
co-added, rejecting cosmic rays. All apart from the $z$-band frames
were photometrically calibrated to the Bessell/Cousins system using
Landolt (1992)\nocite{landolt92} standards. The $z$-band frames were
calibrated to the $z^{\prime}$ filter of the Sloan digital sky survey
(SDSS; York \etal 2000)\nocite{york00}. Atmospheric extinction
corrections of typically 0.05 Magnitudes were applied using archival
data from the ESO ambient conditions database\footnote{{\tt
http://archive.eso.org/asm/ambient-server?night}} for ESO data and
following the prescription of Sung \& Bessell (2000)\nocite{sung00}
for the Taurus images. Corrections for Galactic extinction of
typically A$_{V} = 0.2$ magnitudes were made using values from
NASA/IPAC Extragalactic Database (NED). No corrections were made for
the $z$-band data.

The original intent was to image all fields in all filters. However
due to lack of observing time and instrumental problems this was not
accomplished. Priority was therefore given to those bands that would
maximise the contrast for the 4000\AA \ break.

\subsection{Infrared images}

Near-infrared images were obtained over three runs using the IRAC2B on
the ESO 2.2m telescope and CIRIM on the CTIO Blanco 4m. Fields of view
were smaller than the optical images, $\sim 1^{\prime}$ to $\sim
2^{\prime}$. Pixel scales were \arcsd{0}{22} for CIRIM and
\arcsd{0}{51} for IRAC2B. The seeing was typically
$1^{\prime\prime}$. The observing setups and conditions are again
summarised in Table \ref{tab:runs}.

In all cases, the fields were observed in many short-exposure frames
which were dithered in a non-repeating pattern on the sky. A summary
of the NIR observations is given in Table \ref{tab:obs}. The initial
intention was to obtain NIR data for all objects imaged in the
optical. Due to poor weather, only a subset of these sources were
actually observed in the NIR. These were selected based on RA
observability and on matching optical data.

For the IRAC2B data, the images were dark-subtracted and flat-fielded
with domeflats using programs written in {\tt IDL}. The sky was
removed by making a ``supersky'' from a median stack of the twelve
neighbouring frames in each filter. Photometric calibration was
achieved by observing UKIRT faint standard stars FS11, FS14 and FS27
and faint standards HD56189, HD114895 and HD147778. NED values for
each field were used to correct for Galactic extinction. These were
typically of the order of A$_{\lambda} \sim 0.03$ magnitudes.

Electronic remnants, caused by cross-talk between the readout
amplifiers, are found to occur in some IRAC2B frames. When a bright
object is imaged in one $128 \times 128$ pixel quadrant of the array,
remnants appear an equivalent position in the other three
quadrants. They are typically 5-6 magnitudes fainter than the bright
object and are easily identified by their position and incongruous
colours. All remnants of this type were removed from the catalogues
after detection.

The CIRIM data were reduced using image processing algorithms within
{\tt IRAF\footnote{IRAF is distributed by the National Optical
Astronomy Observatories, which are operated by the Association of
Universities for Research in Astronomy, Inc., under cooperative
agreement with the National Science Foundation.}} and the {\tt IRAF}
package DIMSUM, designed for reducing deep mosaicked images. They first
had the dark current removed by subtracting an average of many dark
frames with the same exposure times as the individual quasar
exposures.  After this, the unregistered frames were combined by
taking an average of all of the exposures of the quasar.  Highly
deviant values were rejected from the average and if bright sources
were visible in the frames, they were masked from consideration during
the combining of all of the frames.  After scaling this flat-field
frame to have an average value of 1. in the middle of the array, the
flat-field was divided into each individual quasar image.  The data
were then run through DIMSUM to register and combine all images of the
same field.

The flux calibration of the images was accomplished using observations
of UKIRT faint standard stars at the beginning and end of each night.

\subsection{Catalogue creation}

\subsubsection{Optical}

For each field, the individual frames in each filter were added to
improve signal-to-noise and catalogues of object positions were
compiled from the resulting deep images using the program SExtractor
\cite{bertin96}. The detection threshold was set in order to avoid a
large number of spurious detections while ensuring that all objects
visible by eye were extracted. These thresholds were 1.5$\sigma$ per
pixel above noise for 4, 8, 5 and 10 contiguous pixels for the EFOSC,
EFOSC2, Taurus and PFCCD data respectively. The resultant positions
were used to calculate aperture magnitudes in each filter using
\arcsd{3}{0} diameter circular apertures. Completeness limits were
estimated by identifying the magnitude at which the number-count
histogram turned over. They are correct to the nearest 0.5 mag and are
indicated in Table~\ref{tab:obs}.

\subsubsection{Infrared}

Because of the much smaller field size of the infrared frames and the
increase in the noise toward the edges of the fields, objects were
extracted from the NIR fields separately. In the case of the IRAC2B
data, where both filters were available, the $J$- and
$K^{\prime}$-band frames in each field were shifted, co-added and
SExtractor was run on this co-added image. The detection threshold was
4 contiguous pixels at 1.5$\sigma$ per pixel above noise.

For the CIRIM $H$-band data, each pixel was split into 4 pixels and
the detection threshold was made 30 contiguous pixels at 2.0$\sigma$
per pixel above noise. This ensured that all objects visible by eye
were detected while spurious sources in the noise to the edges of the
image were minimized.

In both instances, magnitudes were calculated using \arcsd{3}{0}
apertures. Completeness limits were again estimated to the nearest 0.5
mag using the number counts versus magnitude histogram and are
indicated in Table~\ref{tab:obs}.

\section{Single-filter clustering statistics}
\label{sec:sing}

First we quantify the environments of our quasars by calculating
previously-defined statistics based on counting objects within some
radius of the quasar in a single filter. These values can then be
compared directly with other studies.  However, we note that our
observations are not optimised for this analysis due to the absence of
``control'' fields. A critique of these methods is the subject of
\S\ref{sec:sing_crap}. In \S\ref{sec:mult} we utilise the multi-filter
approach for which our observations are best suited.

Here, we have calculated two statistics, the galaxy-quasar
cross-correlation amplitude, $B_{gq}$, and the Hill \& Lilly quantity,
$N_{0.5}$. A brief explanation of the method of calculation employed
in each case is given in Appendix~\ref{app:bgq}. The analysis follows
Wold \etal (2000)\nocite{wold00}, hereafter W00.

\begin{table*}

\begin{minipage}{\linewidth}

\begin{center}

\caption{Physical and clustering properties for our fields. The radio
spectral index, $\alpha$ is measured between 408 MHz and 4.86 GHz,
$\alpha^{4.86}_{0.408}$; $S_{\nu} \propto \nu^{-\alpha}$. D is the
largest radio linear size in kpc. The second column from the right is
an estimate of whether the RLQ belongs to a cluster (C), compact group
(G) or field-like environment (F) based on the multicolour
data. Justification for these estimates are presented in
\S\ref{sec:indi}.}

\begin{tabular}{ccccrr@{\,$\pm$\,}lr@{\,$\pm$\,}lcc}

\hline\hline

MRC quasar & $z$ & 
& $\alpha$ & 
D (kpc)$^{1}$ & \multicolumn{2}{c}{$B_{gq}$ (Mpc$^{1.77}$)$^{2}$} & \multicolumn{2}{c}{$N_{0.5} \, ^{3}$}  & F/G/C & Notes$^{4}$ \\

\hline

B0030--220 & 0.806 & 
& 0.93 & 
$\leq5.8$ & $646$ & $439$ & $6.1$ & $5.3^{*}$ & ? & N \\

B0058--229 & 0.706 & 
& 0.95 & 
503.1 & $-52$ & $362$ & $-1.0$ & $4.5^{*}$ & G & N \\

B0209--237 & 0.680 & 
& 0.92 & 
142.3 & $1377$ & $507$ & $16.0$ & $3.3^{*}$ & C & N \\

B0346--279 & 0.989 &
& 0.17 & 
...  & $1957$ & $1207$ & $10.2$ & $5.7^{*}$ & F$^{5}$ & N \\

B0450--221 & 0.898 & 
& 1.11 & 
120.0 & $335$ & $300$ & $10.7$ & $4.3$ & C & N \\

B0941--200 & 0.715 & 
& 0.88 & 
382.1 & $317$ & $305$ & $2.3$ & $5.6^{*}$ & G & N \\

B1006--299 & 1.064 & 
& 0.94 & 
156.6 & $2092$ & $1304$ & $17.5$ & $6.5^{*}$ & F$^{5}$ & N \\

B1055--242 & 1.090 & 
& 0.51 & 
... & $884$ & $1391$ & $9.4$ & $5.7^{*}$ & F &   \\

B1121--238 & 0.675 & 
& 1.09 & 
362.8 & $175$ & $263$ & $1.9$ & $5.4$ & F \\

B1202--262 & 0.786 & 
& 0.53  & 
122.9 & $470$ & $232$ & $4.2$ & $5.1$ & ? & N \\

B1208--277 & 0.828 & 
& 0.83 & 
359.3 & $582$ & $292$ & $11.0$ & $5.7$ & C & N \\

B1217--209 & 0.814 & 
& 0.94 & 
245.9 & $-11$ & $203$ & $0.5$ & $5.3$ & F \\

B1222--293 & 0.816 & 
& 0.76 & 
242.7 & $-263$ & $229$ & $-9.5$ & $6.4$ & F \\

B1224--262 & 0.768 & 
& 0.83 & 
$<16.3$ & $460$ & $307$ & $8.9$ & $4.0$ & C & N \\

B1226--297 & 0.749 & 
& 0.84 & 
522.7 & $31$ & $260$ & $7.5$ & $6.0$ & F \\

B1247--290 & 0.770 & 
& 0.91 & 
469.7 & $359$ & $325$ & $5.2$ & $5.2$ & F \\

B1301--251 & 0.952 & 
& 0.79 & 
73.6 & \multicolumn{2}{c}{--} & $3.4$ & $6.2^{*}$ & C & N \\

B1303--250 & 0.738 & 
& 0.80 & 
310.9 & $-219$ & $210$ & $8.5$ & $5.0$ & ? & N \\

B1349--265 & 0.934 & 
& 0.63 & 
$<16.9$ & $-420$ & $948$ & $-5.6$ & $6.6^{*}$ & F & \\

B1355--236 & 0.832 & 
& 0.95 & 
132.6 & $31$ & $418$ & $5.9$ & $6.3$ & G/C & N \\

B1359--281 & 0.802 &
& 0.52 & 
11.5 & $518$ & $809$ & $7.6$ & $6.1^{*}$ & F & \\

\hline
\label{tab:bgq}
\end{tabular}

\end{center}

$^{1}$ No value of D is given for unresolved flat-spectrum ($\alpha <
0.52$) objects. These are likely to be highly beamed sources.

$^{2}$ $B_{gq}$ cannot be calculated for MRC~B1301--251 as the $I$-band
image is not deep enough to provide an accurate normalisation of the
luminosity function.

$^{3}$ The values of $N_{0.5}$ indicated with an asterisk are
those where the image was not deep enough to be complete to $m_{g}+
3$. In these cases only objects in the range $m_{g}$ to the
completeness limit were included in the calculation

$^{4}$ Indicates that there is a note on this field in \S\ref{sec:indi}

$^{5}$ The $B_{gq}$ and $N_{0.5}$ values of these RLQs are affected by
what are likely to be foreground systems.

\end{minipage}

\end{table*}

\begin{figure}

  \begin{center}

    \epsfig{file=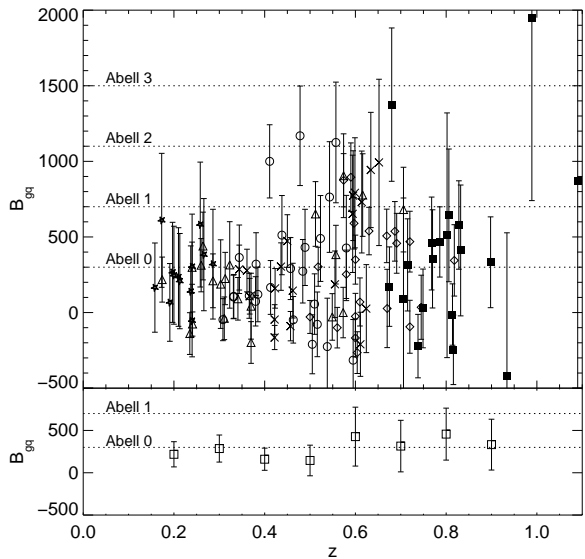,angle=0,width=7.6cm}

    \caption{{\em Upper panel:} Values of the spatial
    cross-correlation amplitude as a function of redshift for
    radio-loud quasar fields. Filled squares: This work; diamonds:
    Wold \etal (2000); triangles: Yee \& Ellingson (1993); circles:
    Ellingson \etal (1991);  crosses: Yee \& Green (1987); stars:
    McLure \& Dunlop (2001). {\em Lower Panel:} The median values of
    $B_{gq}$ from 0.2 to 0.9 at intervals of 0.1 in redshift. The
    horizontal lines in both panels represent the Abell classes quoted
    in McLure \& Dunlop.}

    \label{fig:bgq}

  \end{center}

\end{figure}

\begin{figure}

  \begin{center}

    \epsfig{file=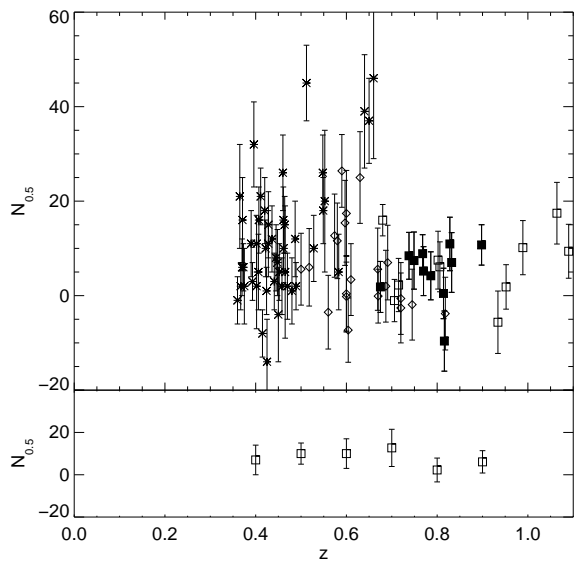,angle=0,width=7.6cm}

    \caption{{\em Upper panel:} Values of the Hill \& Lilly quantity
    {\em vs.} redshift for radio-loud quasars and radio galaxies. Diamonds:
    Wold \etal (2000); asterisks: The Hill \& Lilly (1991) sample of
    radio galaxies; filled squares: This work where $N_{0.5}$ has been
    calculated by counting galaxies to $m_{g} + 3$ at the faint
    end; open squares: This work where $N_{0.5}$ has been calculated
    by counting galaxies to the completeness limit. {\em Lower Panel:}
    The median values of $N_{0.5}$ from 0.4 to 0.9 at intervals of 0.1
    in redshift.}

    \label{fig:n05}

  \end{center}

\end{figure}

The results for the single-filter clustering statistics are shown in
Table~\ref{tab:bgq}.  Our analysis extends these measures to a
redshift of 0.9 with a degree of accuracy comparable to lower-$z$ work
in the literature. Figure~\ref{fig:bgq} and Figure~\ref{fig:n05}
confirm the large variance in $B_{gq}$ and $N_{0.5}$ in the fields of
RLQs as noted by Yee \& Green (1987) \nocite{yee87}. The strength of
clustering around $z \sim 1$ quasars is not significantly different
from clustering around lower-$z$ quasars, implying that there is
little or no evolution in the clustering environments of radio-loud
quasars with redshift.

Above a redshift of 0.9, calculations of $B_{gq}$ have large
uncertainties, and values of $N_{0.5}$ are systematically
underestimated due to lack of depth of the images. No conclusions can
be drawn from single-filter analysis of our sample at $z>0.9$.

The distribution in the richnesses of the environments is found to be
consistent with those found by W00. Our values of $B_{gq}$ and
$N_{0.5}$ were tested against those of W00 using the two-sided KS
test. The values of $B_{gq}$ were restricted to values below $z=0.9$
due to the large uncertainties at greater redshifts. We found $>80\%$
probability that our values of $B_{gq}$ at $z<0.9$ are drawn from the
same population as those of W00. This indicates that there is no
evolution in the richnesses of the environments of RLQs from the
median redshift of the W00 sample ($z \sim 0.6$) to our median
redshift ($z \sim 0.8$).

This consistency is not as evident when testing the values of
$N_{0.5}$. There is a $35\%$ probability of the two samples being
drawn from the same population. However, $\sim 50\%$ of our values of
$N_{0.5}$ cannot be calculated from the full $m_{g} \rightarrow
m_{g}+3$ magnitude range due to the incompleteness of the data. We
also note that the two-sided KS test is not optimised for samples with
significant uncertainties such as this. In order to account for the
uncertainty we ran a Monte Carlo analysis, varying the values of
$N_{0.5}$ within the errors according to a Poissonian distribution and
submitting these values to the KS test. The result was a distribution
of probabilities which was flat, meaning that the W00 sample and our
values of $N_{0.5}$ are as likely to be drawn from a single parent
population as not.

Similar Monte Carlo analyses were carried out on the $B_{gq}$ and
$N_{0.5}$ samples of Yee \& Green (1987), Ellingson \etal (1991), Hill
\& Lilly (1991), Yee \& Ellingson (1993), and McLure \& Dunlop
(2001)\nocite{yee87,ellingson91,hill91,yee93,mclure01}. These yielded
no evidence for different parent populations. We thus conclude that
our calculations of $B_{gq}$ provide no evidence for epoch-dependent
clustering about RLQs.

This finding, as well as those of W00 and Mclure \& Dunlop
(2001)\nocite{mclure01}, is in contrast to those of Yee \& Green
(1987)\nocite{yee87} and Hill \& Lilly (1991) who claimed to detect an
increase in the richness of radio source environments at $z \sim 0.5$.

However, the quasars used in the Yee \& Green investigation were
amalgamated with lower-power RLQs in the study of Ellingson \etal
(1991)\nocite{ellingson91} who concluded that the only the variance
in the clustering richness, rather than $B_{gq}$ itself, increased
with $z$. W00 also found no evidence for redshift-dependent evolution
in environment based on analysis of $B_{gq}$ using the Spearman
statistic. Instead, by disentangling radio power and redshift
information, it was calculated that the source of the epoch-dependent
clustering signal was a weak radio power-$B_{gq}$ relationship
combined with the luminosity-redshift relationship inherent in the
sample selection function. Hill \& Lilly found that the environments
of their radio galaxies were indistinguishable from the quasars of Yee
\& Green. The radio power-clustering relationship can therefore also
be used to explain their findings. FRII radio galaxies in Hill \&
Lilly's $z \sim 0.5$ sample are, on average, $>10$ times more powerful
than those at $z \sim 0$.

More prosaic factors may also contribute to the discrepancy between
the Yee \& Green (1987)\nocite{yee87} and Hill \& Lilly (1991) studies
and more recent results. The $1-3$m telescopes used by the former
investigators were operating near their detection limits at $z \sim
0.5$ making determination of the clustering statistics at these
redshifts uncertain. Furthermore, the small number of fields at the
highest redshifts combined with the failure of $B_{gq}$ and $N_{0.5}$
in individual cases above $z \sim 0.4$ (as discussed in
\S\ref{sec:sing_crap}) could have served to distort the statistics.

\section{Clustering of red galaxies in the fields of RLQs}
\label{sec:mult}

Because we know the redshift of the RLQ, we can select optical filters
which isolate the redshifted 4000\AA \ break in core cluster
ellipticals. This ensures that elliptical galaxies at the RLQ redshift
appear not only as the reddest objects in the field, but are also rare
when compared with the colours of all other galaxies in the field. It
is also well established that the colours of elliptical galaxies in
the cores of clusters fall in a narrow and predictable
colour-magnitude range (\eg Baum 1959; Bower \etal 1992; Gladders
\etal 1998; Stanford \etal
1998\nocite{baum59,bower92,gladders98,stanford98}). These factors
serve to increase the ``contrast'' of cluster cores with respect to
their background in our images and so make their presence more
detectable.

The NIR data have a small field of view, precluding us from examining
an area further than $\sim 2^{\prime}$ from the quasar. Since the
images are relatively shallow, we will detect no more than the
brightest galaxies in any putative cluster. The strength of these data
come in the identification of individual galaxies. The $R-K$ and $I-K$
colours of elliptical galaxies follow a highly predictable path with
redshift as shown in Figure~\ref{fig:coltab}. The $J-K$ colour is
important as it resolves two degeneracies not evident through the sole
use of $R-K$ or $I-K$ colours. It sets unresolved high-redshift
galaxies completely apart from faint stars and brown dwarfs which have
$J-K<1.5$ and also distinguishes old elliptical galaxies whose colours
are $1.5<J-K<2.0$ at $0.6<z<1.1$ from dusty star-forming galaxies
which have colours $J-K>2$ at the same redshifts \cite{pozzetti00}.

More generally, NIR colours probe the old stellar population and are
less prone to change influenced by short bursts of star formation. We
would therefore expect to be able to identify the objects in suspected
clusters with cluster core members with a greater degree of accuracy
using additional $J$, $H$ and $K'$ filters than if we relied solely on
optical colours. Our images can be compared with surveys observed in
NIR, \eg the Herschel Deep Field \cite{mccracken00}, in order to
quantify the size of any general overdensities.

\subsection{Red galaxies in optical filters}

\begin{figure*}

  \hbox{
  \epsfig{file=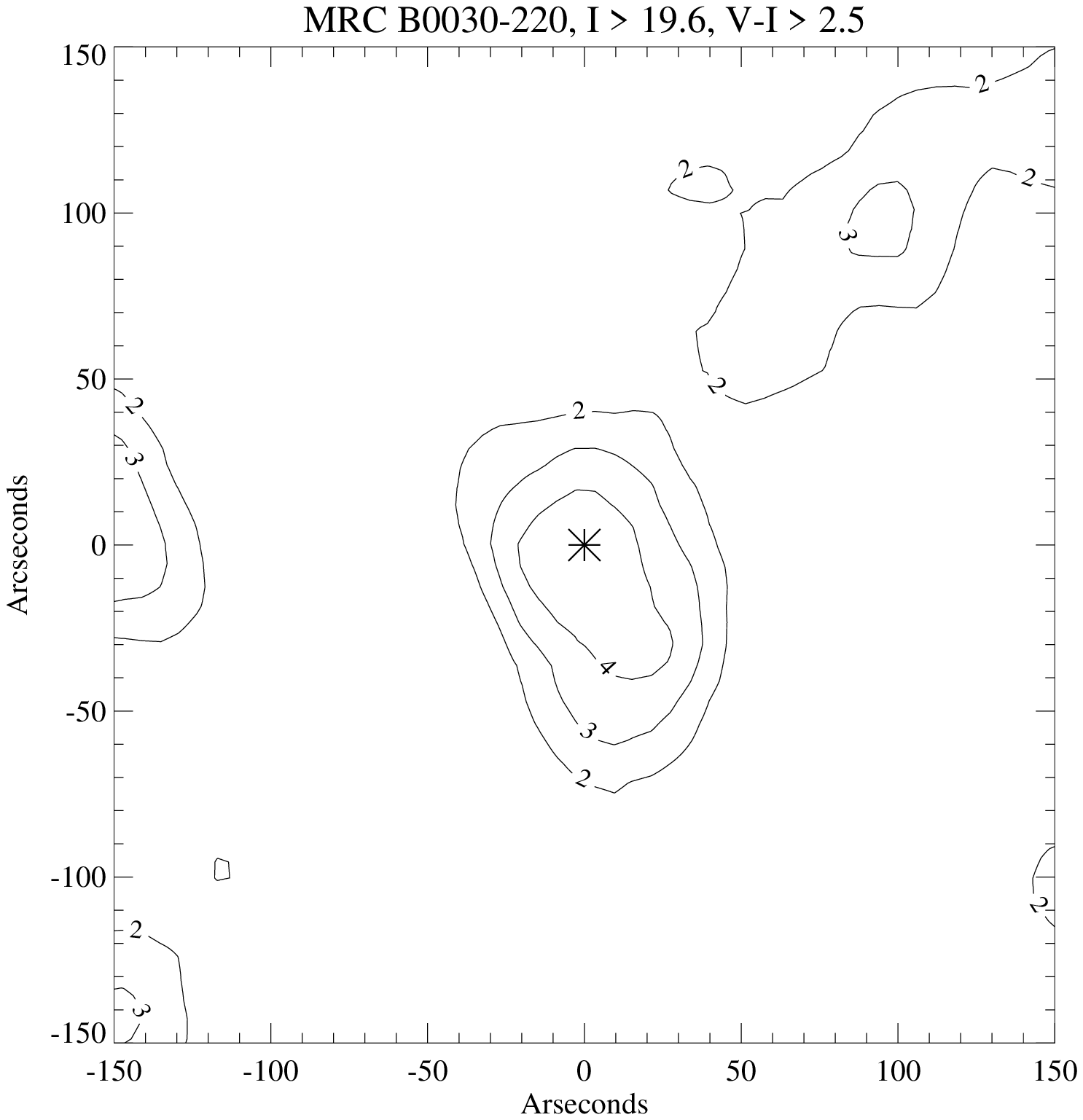,angle=0,width=5.5cm}

  \epsfig{file=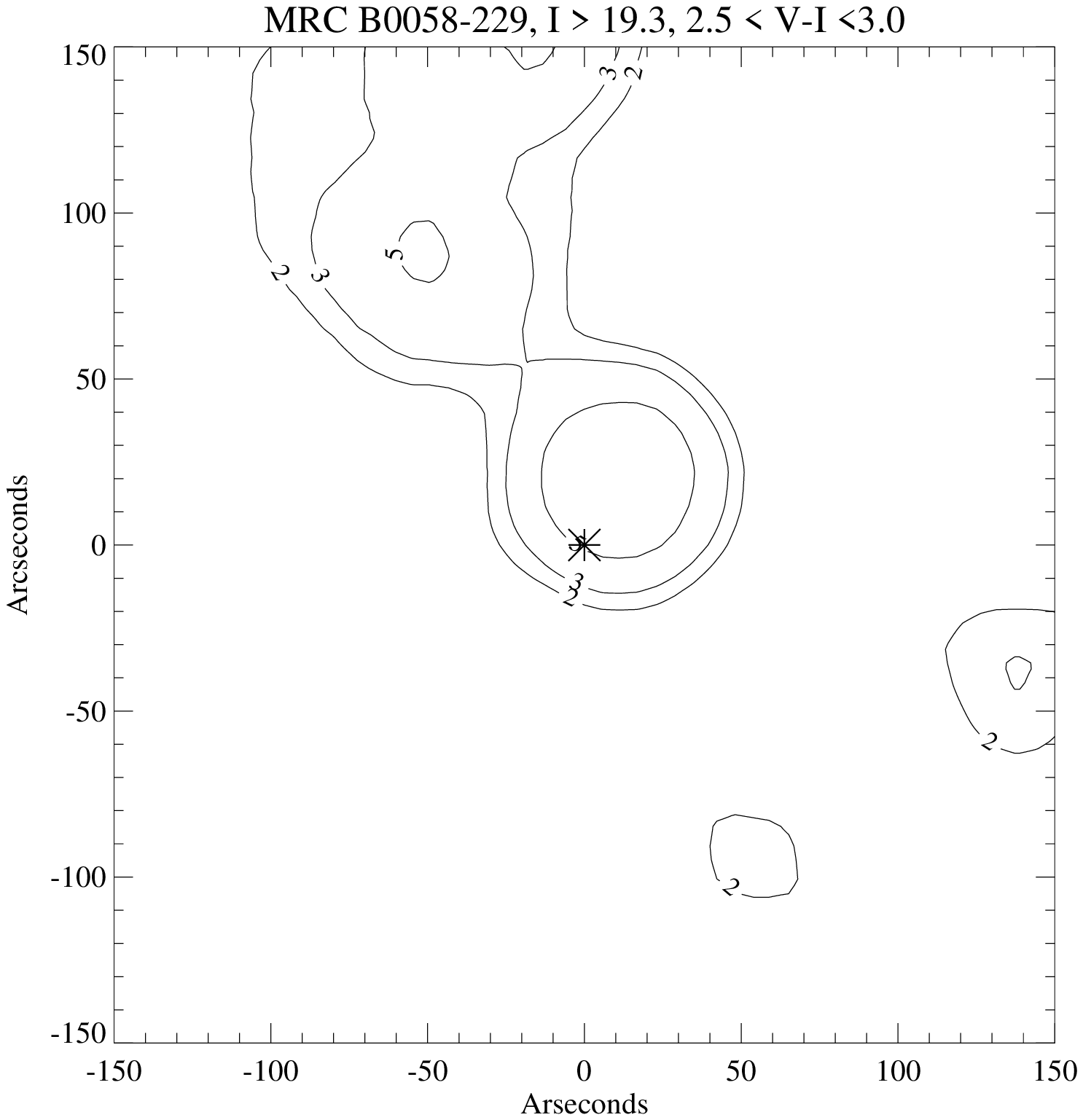,angle=0,width=5.5cm}

  \epsfig{file=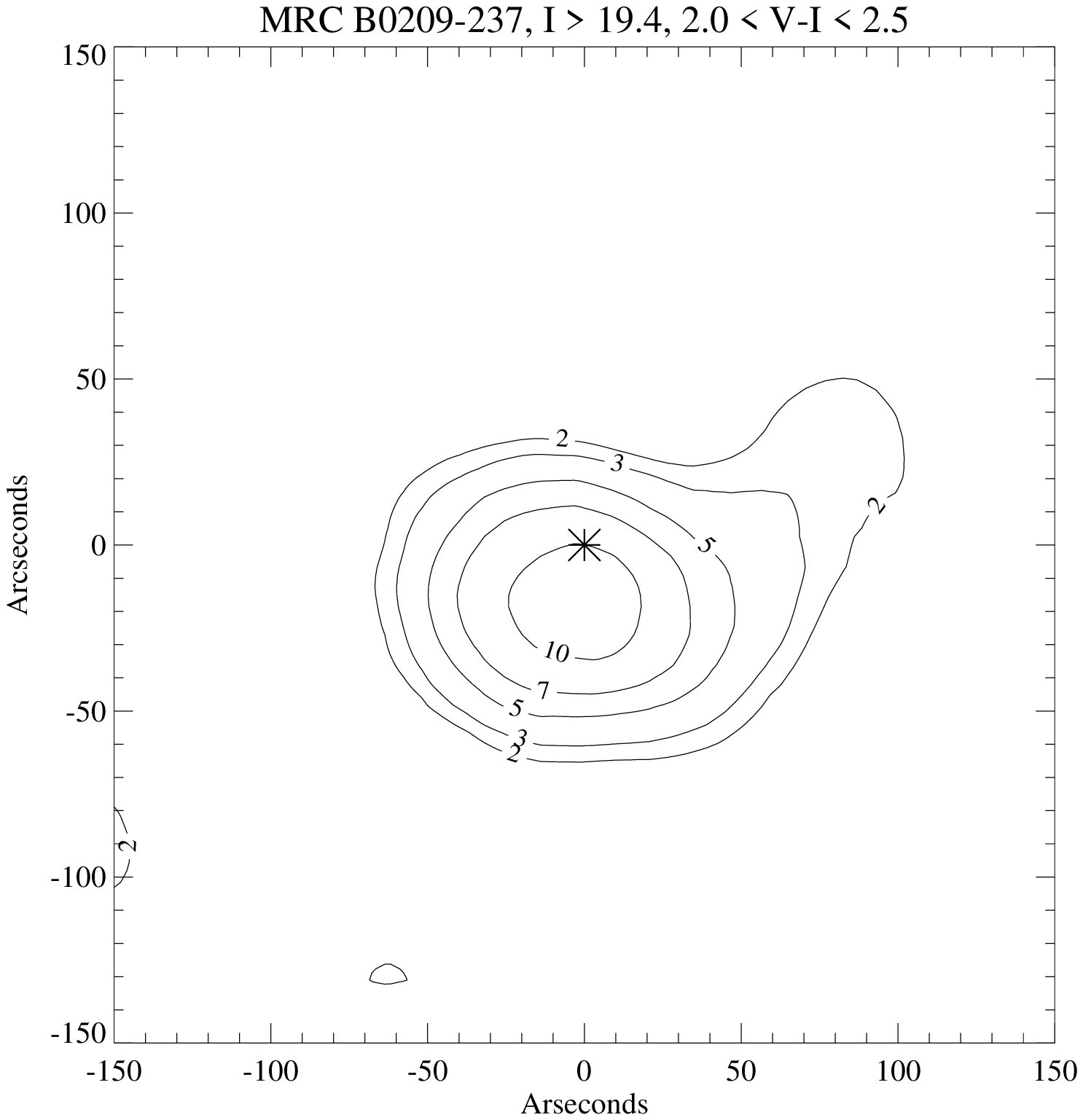,angle=0,width=5.5cm}
  }

  \hbox{
  \epsfig{file=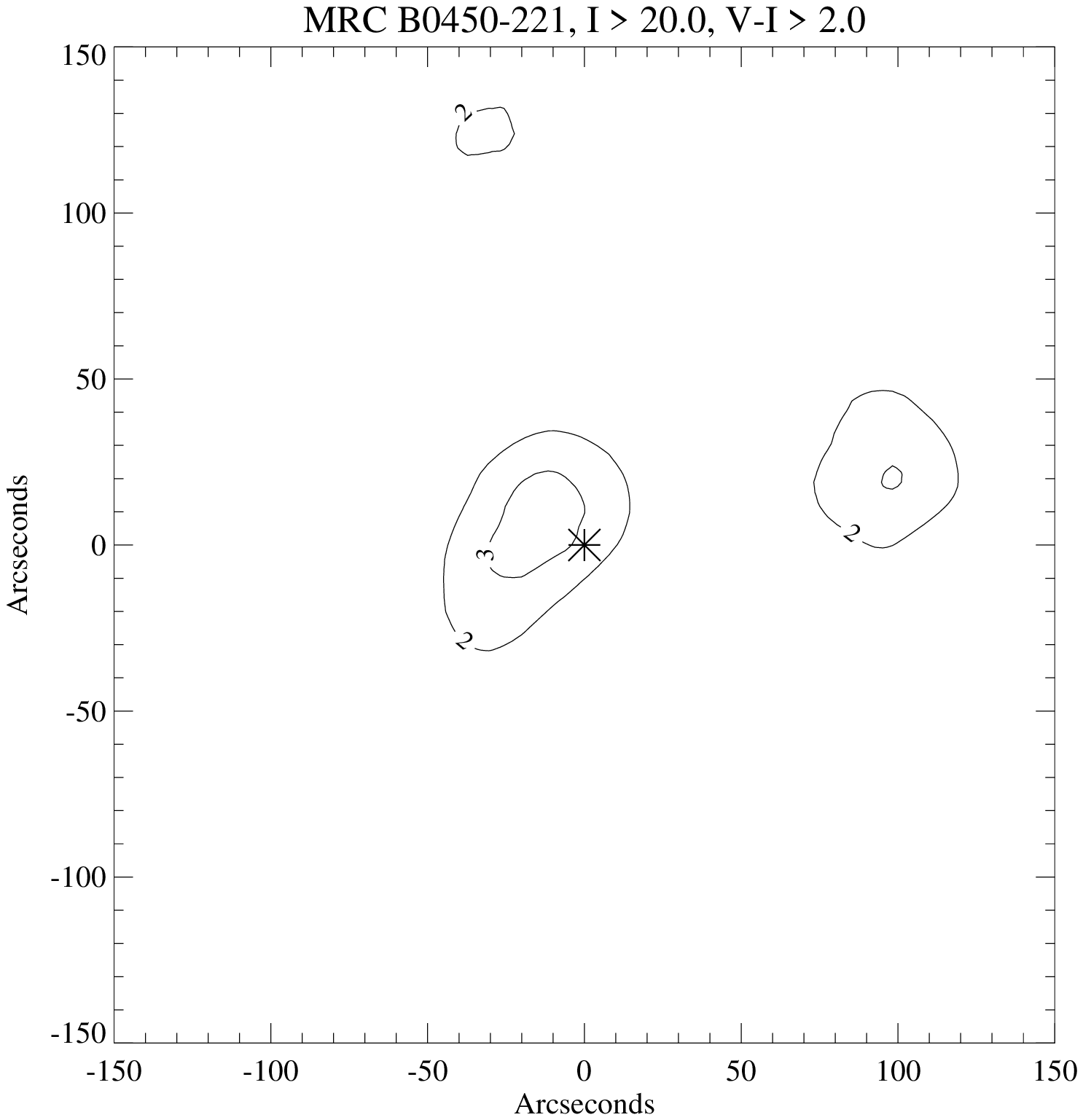,angle=0,width=5.5cm}

  \epsfig{file=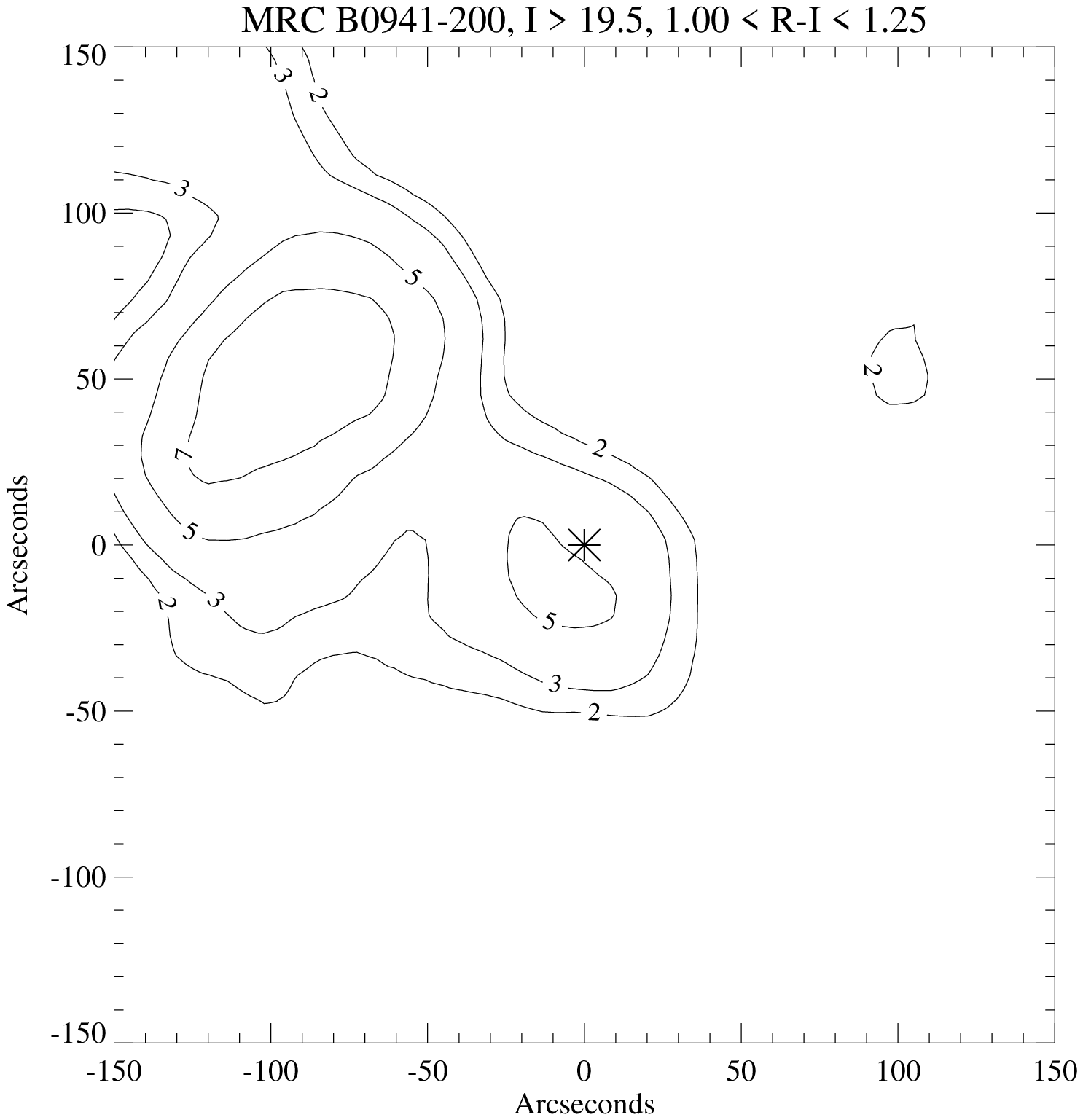,angle=0,width=5.5cm}

  \epsfig{file=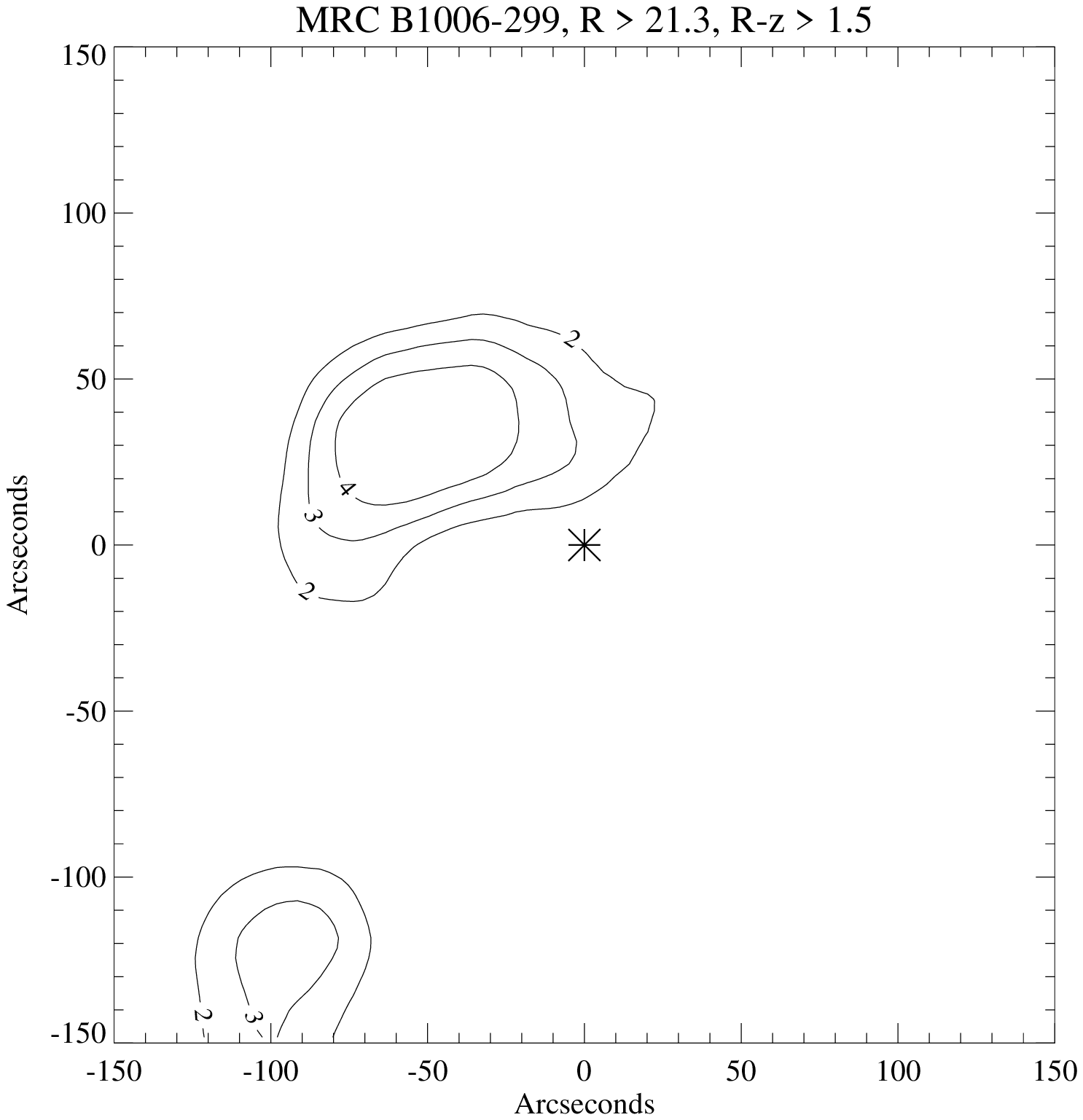,angle=0,width=5.5cm}
  }

  \hbox{
  \epsfig{file=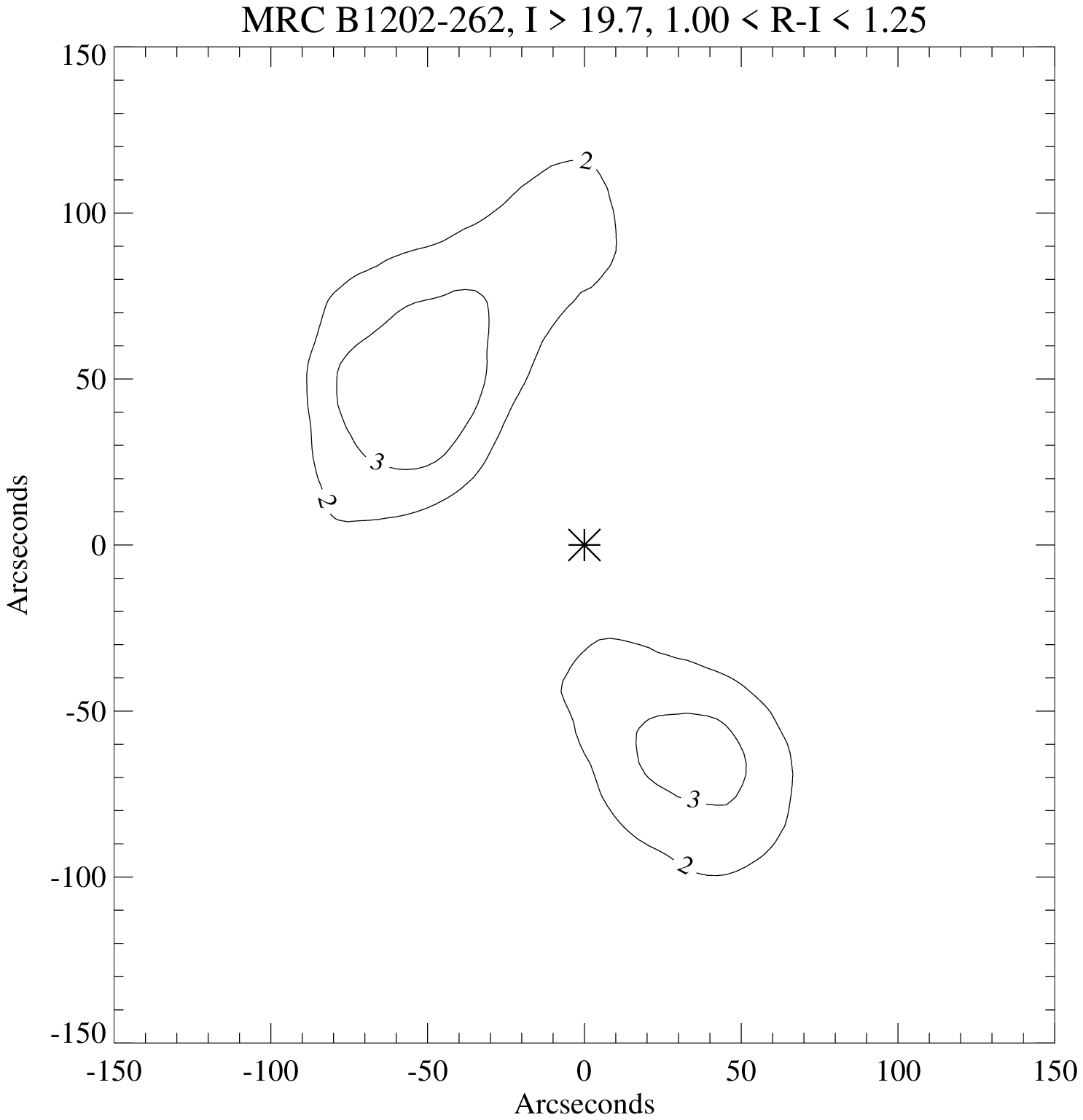,angle=0,width=5.5cm}

  \epsfig{file=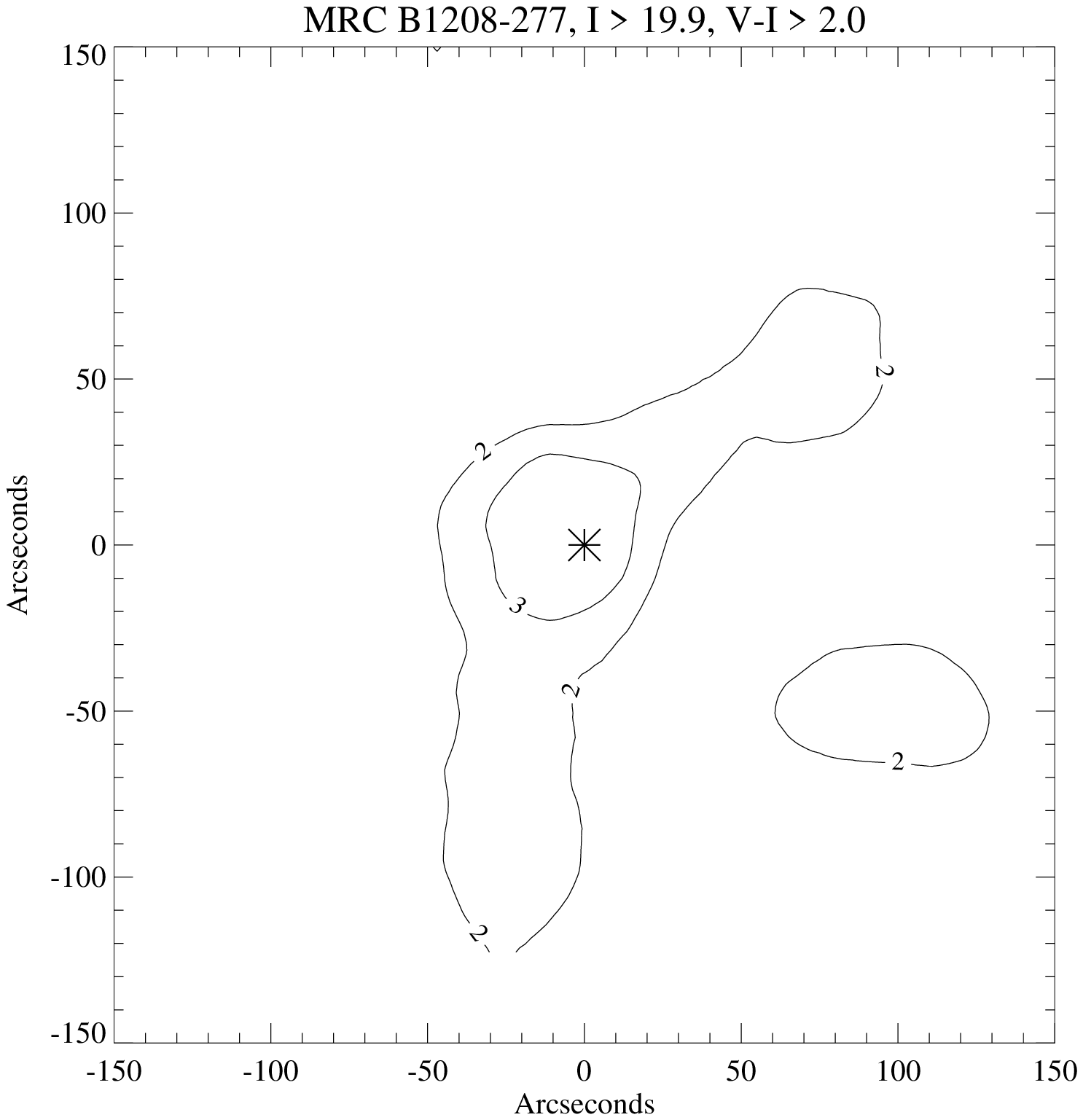,angle=0,width=5.5cm}
 
  \epsfig{file=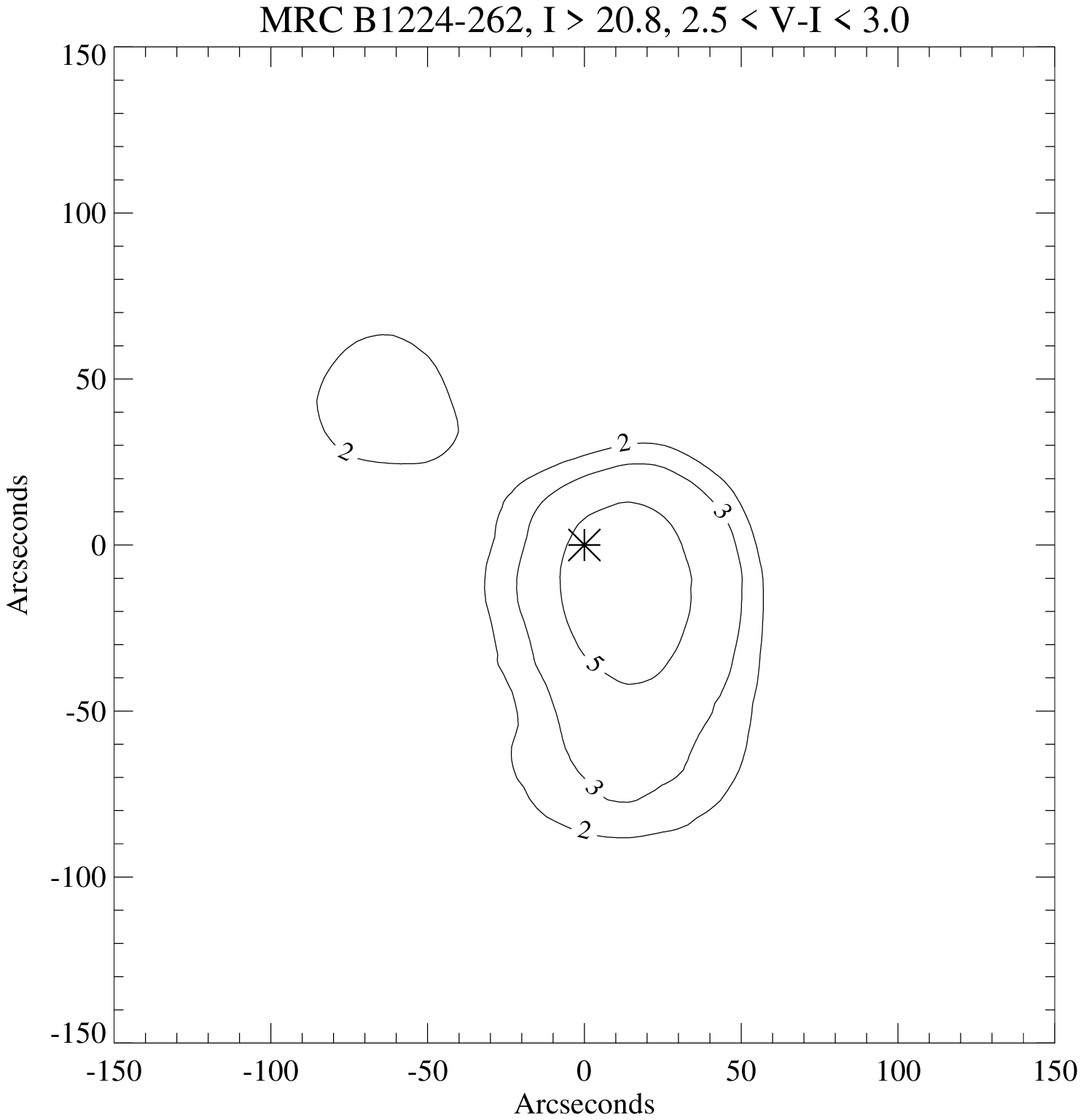,angle=0,width=5.5cm}
  }

  \hbox{
  \epsfig{file=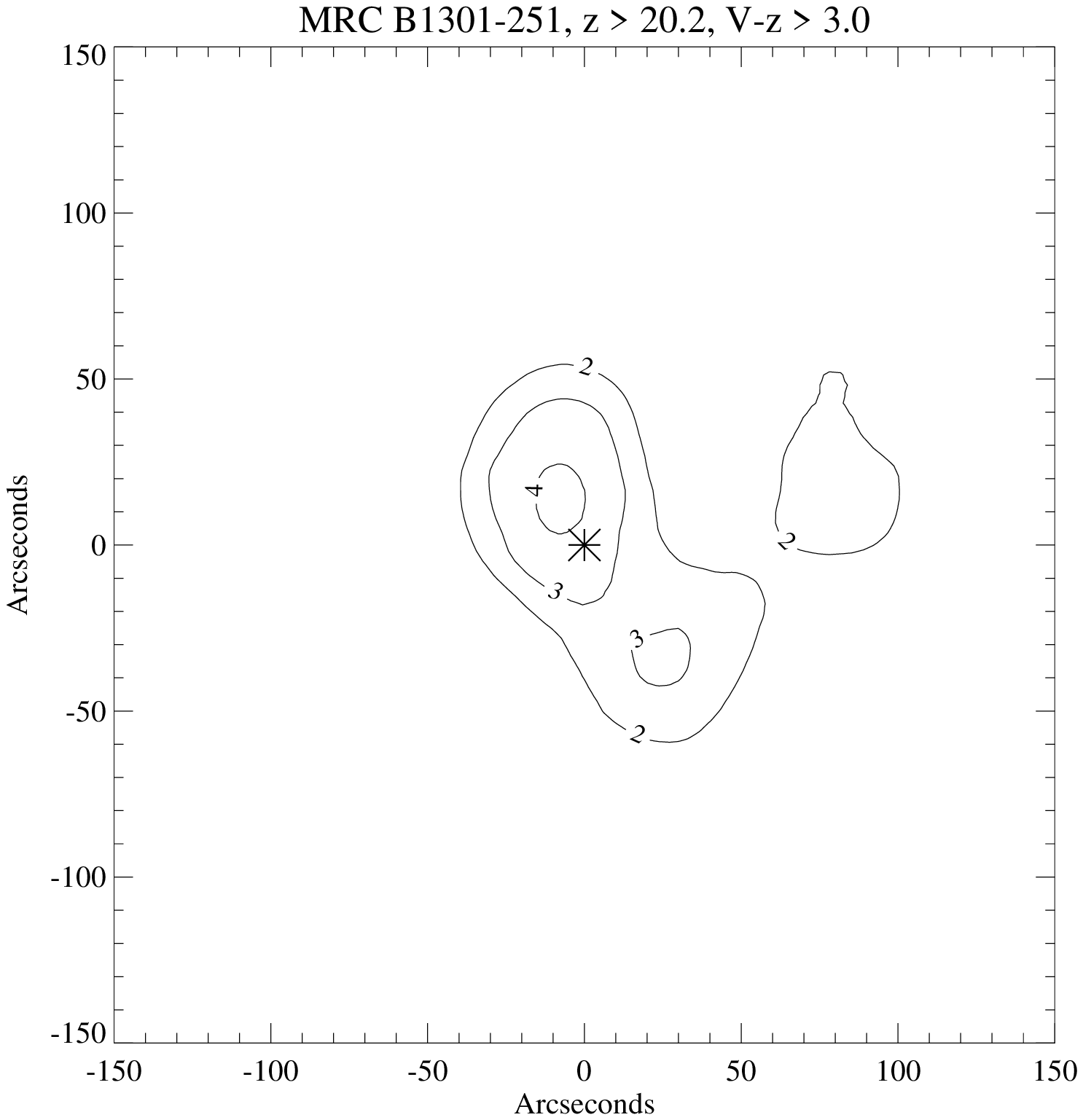,angle=0,width=5.5cm}

  \epsfig{file=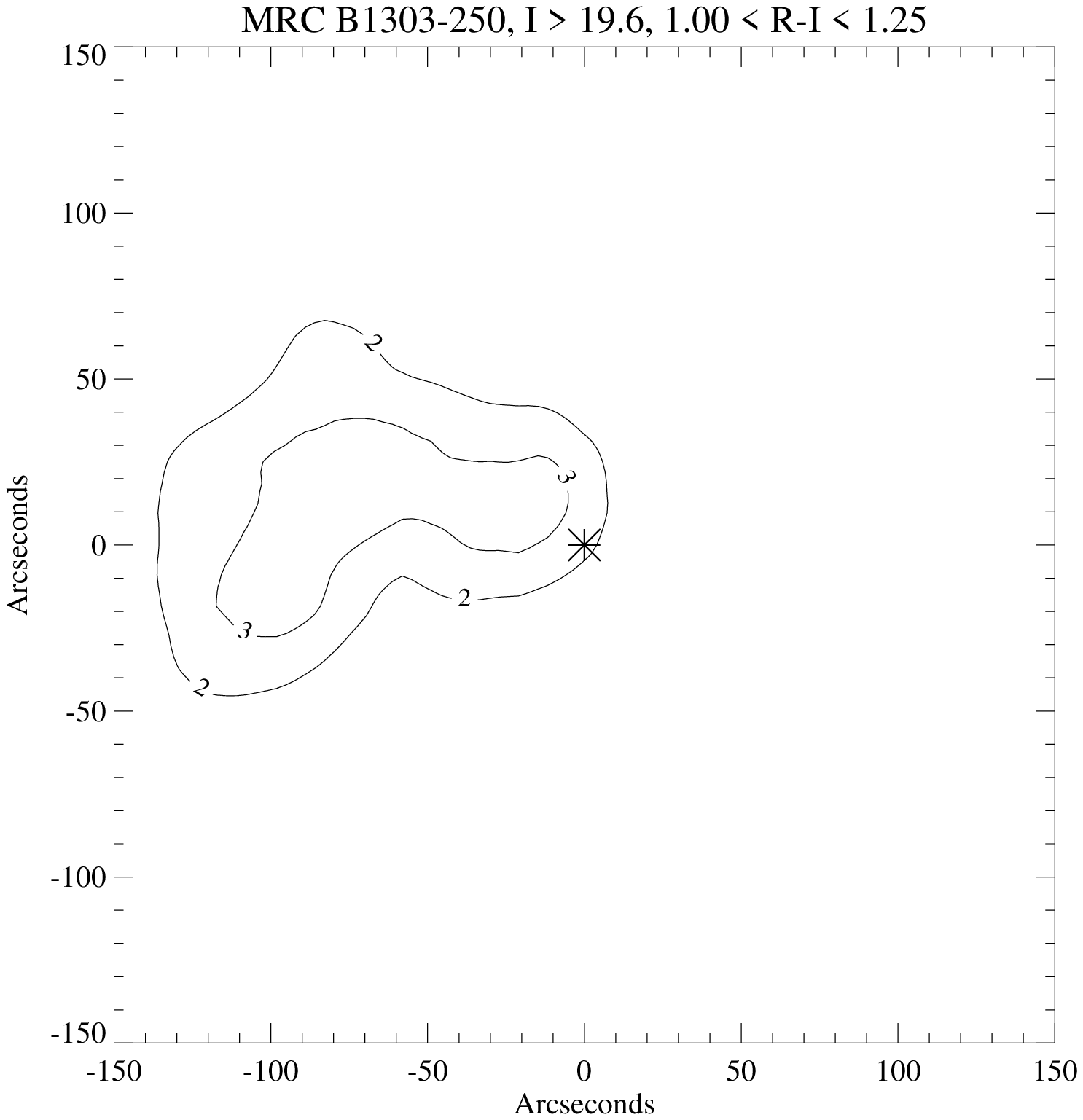,angle=0,width=5.5cm}

  \epsfig{file=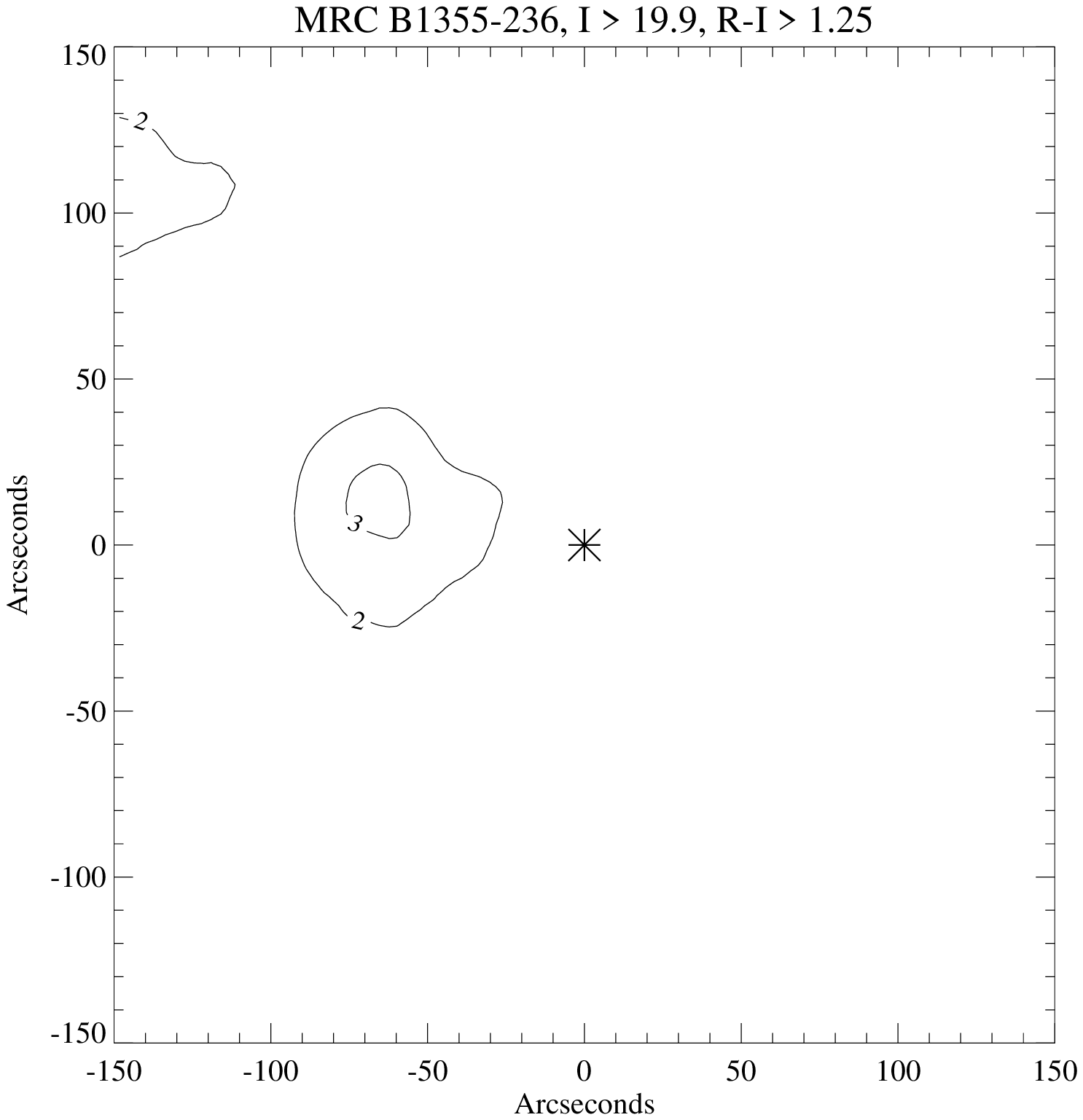,angle=0,width=5.5cm}

  }

  \caption{Diagrams showing the surface density of objects in selected
  fields with the colours of elliptical galaxies at the quasar
  redshift. Contours are in units of $\sigma \times$ the Poissonian
  noise of the background distribution (see text). The RLQ is marked
  with an asterisk. North is up, East is left. The maps were
  constructed using each image's full field of view and then cut to
  $300^{\prime\prime}\times 300^{\prime\prime}$ about the RLQ.}

  \label{fig:grey}

\end{figure*}

Because of the complex nature of the data set, the use of differing
facilities and filters, and the varying depths of our images, there is
no one test which we can employ to find groups or clusters of
galaxies. Multiple techniques must be used to establish the presence
of clustering.

A useful initial search method is to construct maps of the
surface-density distribution of objects. This census is restricted to
those objects with the colours of passively-evolving elliptical
galaxies at the redshifts of the quasars. The panels in Figure
\ref{fig:grey} are examples of this. The surface density is
calculated by counting the number of objects within 300~kpc
($35^{\prime\prime} - 40^{\prime\prime}$ at the redshifts of the
sample -- about the size of a cluster core), of each point in the
field. Only objects fainter than a brightest cluster member (\eg Eales
1985; Snellen \etal 1996; Arag{\'o}n-Salamanca, Baugh \& Kauffmann
1998)\nocite{eales85,snellen96,aragon-salamanca98} and with the
colours of core cluster ellipticals at the relevant redshift are
considered. This number is then normalised by the RMS uncertainty in
the rest of the field (\ie outside the 300~kpc counting radius),
assuming a Poissonian noise distribution. Note that this does not give
uniform results from field to field as it inevitably depends on local
field richness, as well as the colours of the objects targetted.

Each map is boxcar-smoothed using a width of 300 kpc to remove
small-scale pixel-to-pixel variations. The widths of the colour bins
are not uniform, but are made such that they are not less than the
errors in the colours of the objects of interest. Above a redshift of
0.8 colour limits are used, so all objects redder than a certain
colour are considered as candidate cluster members. This is to include
those objects which may not be observed in the bluer bands because
their magnitudes are near or below the completeness limit of the
data. Where the $B$ or $V$ band is involved, the width of the colour
bin is larger to allow for the possible brightening of objects due to
small amounts of on-going star formation in cluster ellipticals.

Each frame in Figure~\ref{fig:grey} shows the density contours of
objects in each field with the colours of passively-evolving
elliptical galaxies at the redshift of the quasar. As these colours
are both filter and redshift dependent, their value and range varies
according to the specific observation.  To correct for the differing
fields of view, the maps were calculated over each full image and then
cut to $300^{\prime\prime}\times 300^{\prime\prime}$ about the RLQ.

Simulated maps were created to analyse how often chance superpositions
of objects would cause false signals. These were assembled by
distributing `objects' randomly in $x,y$ space. Total numbers of these
artificial data were chosen to replicate the numbers detected in the
real data. The surface-density calculations were then made to produce
maps like those in Figure~\ref{fig:grey}. Peaks of $>5\sigma$ were
seen in $\lesssim 0.05\%$ of the simulated fields suggesting that
those seen in Figure~\ref{fig:grey} represent physically associated
systems. Three-sigma peaks, in contrast, were common enough to be
confused with real clusters of galaxies (Occurring at least once in
nearly all of the fields). Maps with $3 - 4\sigma$ peaks near the
quasar were therefore examined for some other evidence (\eg NIR
colours) to be included as cluster candidates.

\begin{figure*}

  \begin{center}

    \epsfig{file=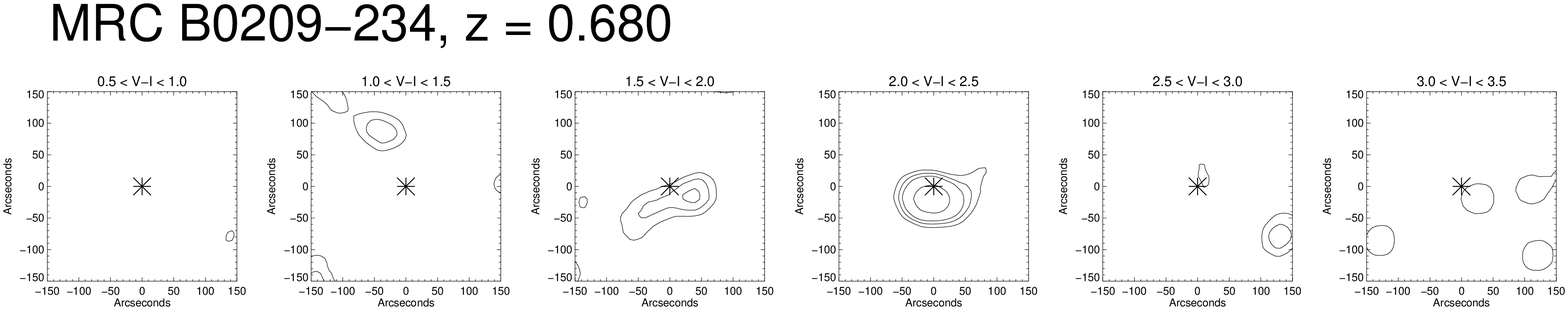,angle=0,width=16cm}

  \caption{The density enhancements of objects of different colours in
  the field of MRC~B0209--279. The contours represent units of 2, 3,
  5, 7 and 10 $\times$ the Poisson noise of the
  background. Passively-evolving ellipticals at $z=0.680$ are expected
  to appear in the $2.0<V-I<2.5$ panel.}

  \label{fig:0209_cont}

  \end{center}

\end{figure*}

We can also construct these maps over a range of colour, \eg
Figure~\ref{fig:0209_cont}. The surface density is
calculated for objects in different colour bins. Each point in the
field is therefore given several (colour dependent) values for the
clustering. In this way, a field with a high value of $B_{gq}$ can be
examined to see if the colours of the clustering galaxies are
consistent with an overdensity about the quasar.

We use the surface density of red galaxies as an initial method to
identify cluster candidates. Unlike the measurements used by previous
investigators, it does not rely on the centering of a particular
statistic on the AGN. In fact one of the most interesting things that
we see initially is that the RLQ is not always to be found at the
centroid of the red galaxy distribution. It is also possible to
distinguish an agglomeration at the redshift of the quasar from one
which isn't, (\eg see MRC~B0346--279 and MRC~B1006--299 in
\S\ref{sec:indi}). Because of the relative paucity of redder objects
relative to those of other colours, this method is optimised for
finding groups and clusters of galaxies at high redshift.

We detect $>5\sigma$ overdensities with the correct colours in four
fields. Overdensities of $3<\sigma<5$ are found within 0.5 Mpc of the
quasar in a further eight fields. These latter objects are examined
for further evidence that the clustering galaxies are at the redshift
of the quasar. Examinations of the fields individually are the subject
of \S\ref{sec:indi}.

\subsection{Infrared imaging results}

\begin{figure*}

  \begin{center}

  \epsfig{file=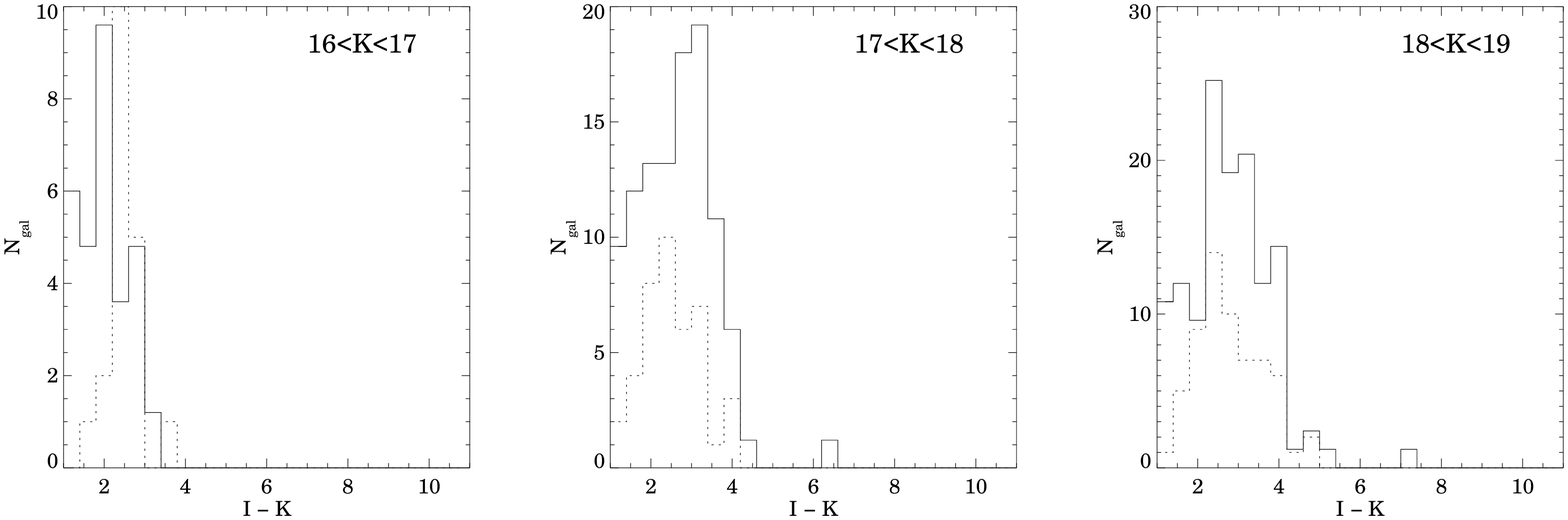,angle=0,width=17cm}

  \caption{Comparison of the $I-K^{\prime}$ number counts for fields
  in our sample imaged in the $K^{\prime}$ (solid line) with the
  Herschel deep field (dotted line). The Numbers in our sample have
  been normalised to the area of the WHDF but no correction has been
  made for the incompleteness of our sample. At fainter magnitudes we
  see a clear excess of objects around our quasars.}

  \label{fig:whc}

  \end{center}

\end{figure*}

The fields for which we have $K^{\prime}$-band data comprise a total
area of \arcmn{39}{4}. This allows us to make a comparison of our
fields with that of the \arcmn{47}{2} $K$-band selected survey of the
Herschel deep field \cite{mccracken00}. Figure~\ref{fig:whc} shows
this comparison. Our fields show a $2.6 \pm 0.6$ overdensity of objects
with $17<K^{\prime}<19$ and $I-K^{\prime}>3$ over the Herschel deep
field.  Our fields are highly incomplete above $K^{\prime}=18$, so the
true overdensity is probably greater. This strongly suggests that
objects with the colours of passive elliptical galaxies tend to
cluster in the fields of RLQs.

Unfortunately, the $K^{\prime}$ imaging does not constitute a complete
subset of our data. We show in Section~\ref{sec:indi} that five out of
eight of the objects in our $K^{\prime}$ sample have overdensities in
their fields. Thus to draw conclusions regarding clustering about all
RLQs from this subsample is somewhat spurious. In order to say
something about the global properties of the sample, we must examine
the fields individually. In this context, the NIR data is used to
check and support cluster identifications.

\subsection{Notes on selected fields}
\label{sec:indi}

All quasar fields have been examined for clustering of red
galaxies. In this section we detail those sources for which there is
evidence of groups and clusters in their environs. This is either
because of a significant value of $B_{gq}$ or a $>5\sigma$ overdensity
with the appropriate colour within $\sim 0.5$ Mpc of the quasar in
Figure~\ref{fig:grey}. Fields with $3 - 4\sigma$ overdensities within
0.5 Mpc of the quasar are not automatically classified as residing in
clustered environments. Rather, only if there is additional evidence
(\eg an overdensity with the appropriate NIR colours) is a judgement
made. In the cases of MRC~B0030--220, MRC~B1202--262 and
MRC~B1303--250 no determination can be made on the nature of $3 -
4\sigma$ overdensities within 0.5 Mpc of the quasars. Fields not
commented on in this section have been examined and have environments
consistent with those of galaxies in the field, \ie the clustering
about the AGN is consistent with the clustering about any other galaxy
in its field.

Compact groups can be distinguished from clusters by modifying
Hickson's (1982)\nocite{hickson82} criteria for compact groups
according to the prescription of Bremer \etal
(2002)\nocite{bremer02}. Two of Hickson's requirements are
retained. Firstly that there are more than 4 galaxies in the group
whose magnitudes differ by less than 3.0. Secondly, the average group
surface brightness (after adjustment for cosmological dimming and
K-correction) is $\mu < 26$ Mag arcsec$^{-2}$. Bremer \etal note that
the criterion that there be no other galaxies within 3 group radii
brighter than 3 magnitudes fainter than the brightest group member is
extremely difficult to fulfill at high redshifts. This is because of
the increased number counts at these faint magnitude levels. Instead
of this restriction we impose a colour requirement. Namely that there
be no other galaxies within 3 group radii with the same colours as the
group galaxies. This serves to exclude the cores of clusters of
galaxies being classified as groups, which was the original reason for
Hickson's third criterion.

\subsubsection{MRC~B0030--220, $z=0.806$}

The value of $B_{gq}$, $646 \pm 439$, and the $4\sigma$ peak in the
surface density of objects with $V-I>2.5$ in Figure~\ref{fig:grey}
suggest a marginal overdensity. The quasar resides to the north of a
small group of red galaxies ($V-I>2.5$) centred on a galaxy with
$I=20.7$, $V-I=2.9$.

The small significance of the detection is due, in part, to the
richness of the field. SExtractor identified $\sim 1000$ objects in
the \arcmn{8}{5} field, about twice as many as detected in the fields
of other RLQs imaged during the same run. Deeper optical/NIR imaging
or MOS is needed to quantify the exact nature of the clustering. For
the present time, we defer classification of this system.

\subsubsection{MRC~B0058--229, $z=0.706$}

\begin{figure}

  \begin{center}

    \psfig{file=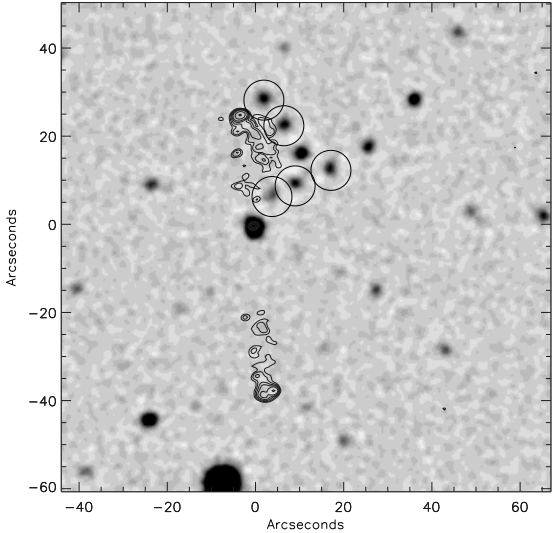,angle=0,width=8cm}

    \caption{The $I$-band image of MRC B0058--229 with the 4860 MHz
    VLA radio contours overlaid. The quasar is located at 0,0; North
    is up, East is left. Galaxies with $20<I<22$ and $2.6<V-I<2.8$ are
    circled. These appear closer to the shorter, northern radio lobe.}

    \label{fig:c0058}

  \end{center}

\end{figure}

There is a $5\sigma$ peak in Figure~\ref{fig:grey} cospatial with a
small group of red objects to the north of the quasar. On closer
inspection, five of these objects are found to have a very low
magnitude and colour dispersion ($20<I<22$; $2.6<V-I<2.8$) consistent
with their being ellipticals at $z=0.706$ and strongly suggesting they
are associated. These 5 objects satisfy the modified Hickson criteria
to be classified as a compact group. Indeed over the rest of the
\arcmn{8}{6} diameter circular field there are only eight other
objects with these colours. The northern radio lobe is shorter by
about $10^{\prime\prime}$ indicating that it may be interacting with
the galaxies or a higher ambient density there
(Figure~\ref{fig:c0058}).

The combined optical and radio evidence suggests that MRC B0058--229
resides at the edge a compact group of galaxies at $z=0.7$.

\subsubsection{MRC~B0209--237, $z=0.680$}

\begin{figure}

  \begin{center}

    \psfig{file=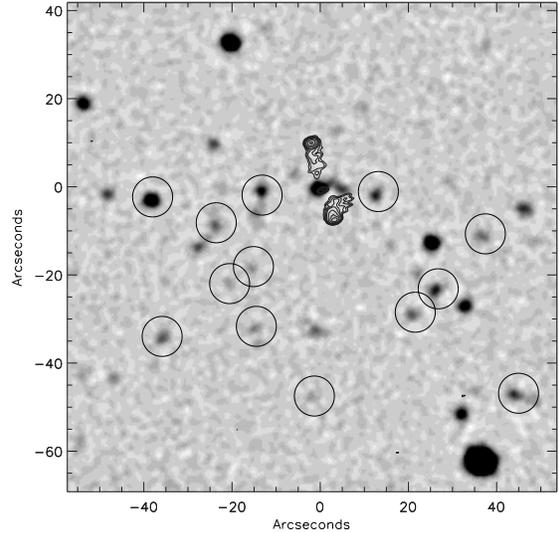,angle=0,width=8cm}

    \caption{The $I$-band image of MRC~B0209--237 overlaid with the
    4860 MHz VLA radio contours. The quasar is at 0,0; North is up,
    East is left. Galaxies with $20<I<23$ and $2.0<V-I<2.5$ are
    circled.}

    \label{fig:c0209}

  \end{center}

\end{figure}

The value of $B_{gq}$ ($1377 \pm 507$) and the $>10\sigma$ overdensity
in Figure~\ref{fig:grey} suggest the presence of a rich cluster of
galaxies. There are 13 objects with $20<I<23$ and $2.0<V-I<2.5$ in the
0.5 Mpc circle centred on the quasar compared with 32 in the rest of
the field. This represents a density of $3.7$ arcmin$^{-2}$ near the
AGN as opposed to $0.6$ arcmin$^{-2}$ in the field, a six-fold
overdensity. The quasar is not found at the centre of the galaxy
distribution, the majority of cluster galaxies being to the south of
the quasar. The radio contours show that the southern radio lobe is
shorter than the northern one, indicative of a denser ICM to the south
(Figure~\ref{fig:c0209}).

\subsubsection{MRC~B0346--279, $z=0.989$}

\begin{figure*}

  \begin{center}

  \epsfig{file=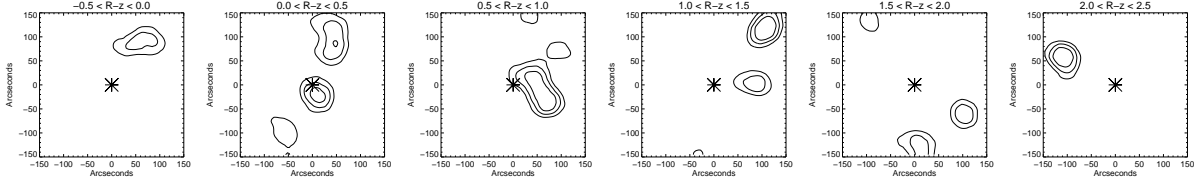,angle=0,width=16cm}

  \caption{The density enhancements of objects of different colours in
  the field of MRC~B0346--279. The contours represent units of 2, 3
  and 5 $\times$ the Poisson noise of the background. The position of
  the quasar is marked by the asterisk. Passively-evolving ellipticals
  at $z=0.989$ are expected to appear in the $1.5<R-z<2.0$ panel.}

  \label{fig:0346_cont}

  \end{center}

\end{figure*}

The quasar has large values of $B_{gq}$ and $N_{0.5}$ ($1957 \pm 1207$
and $16.0 \pm 3.3$) and there are a large number of faint objects
($23.0<R<24.0$) to its west. However, no overdensity is seen in
Figure~\ref{fig:grey}. On closer inspection, the colours of the
objects which contribute to the large values of $B_{gq}$ and $N_{0.5}$
are too blue to be ellipticals at
$z=0.989$. Figure~\ref{fig:0346_cont} shows the surface density of
galaxies in different colour slices. Redshift 1 ellipticals are
expected to have $1.5<R-z<2.0$. The overdensity in the field of
MRC~B0346--279 is caused by objects with $0.0 < R - z < 1.0$, about 1
magnitude bluer. Therefore, this is more likely a lower-$z$
overdensity.

Only 10 objects are detected in NIR of which 3 have $R-K^{\prime} >
5$. The completeness limit of the NIR data is $K^{\prime}= 19.0$,
giving most objects an upper limit of $R-K^{\prime} \sim 4-5$. For a
$z = 0.989$ cluster of galaxies we should expect to see objects with
$18 < K^{\prime} < 19$ and $R-K^{\prime} > 5$. The median colours for
objects with $18 < K^{\prime} < 19$ in this field are $R-K^{\prime}
\sim 3$ and $J - K^{\prime} \sim 1$
(Figure~\ref{fig:0346_colmag}). These colours indicate relatively
low-redshift galaxies (Figure~\ref{fig:coltab}).

If the objects around MRC~B0346--279 were to comprise a cluster it
would be highly unusual as the system has no discernible red galaxy
population. This would be unique, even among the few high-redshift
clusters known to date. Deep multi-object spectroscopy of this field
will be required to resolve these issues. At present, we classify
MRC~B0346--279 as residing in an environment indistinguishable from the
field with a concentration of objects in the foreground.

\subsubsection{MRC~B0450--221, $z=0.898$}

A cluster of galaxies has been detected about this quasar. There is an
overdensity of galaxies with the optical -- NIR colours of 0.898
elliptical galaxies and an enhancement of emission-line galaxies
(ELGs). Preliminary spectroscopy confirms that the ELGs are at the
redshift of the quasar \cite{baker01}.

The ambiguous value of $B_{gq}$ ($335 \pm 300$) for MRC~B0450--221
supports the use of the multicolour detection algorithm. Despite the
relative richness of the cluster, $B_{gq}$ is hardly significant as a
pure overdensity and would not have alerted investigators to the
presence of clustering. The richness of the field further than 0.5 Mpc
from the quasar contributes to the dilution of $B_{gq}$. As well as
this, the high redshift of the source allows only a fraction of its
galaxy population to be visible above the completeness limit.

\subsubsection{MRC~B0941--200, $z=0.715$}

Optical and NIR observations of this field have been reported in
Bremer \etal (2002)\nocite{bremer02}. They discovered evidence for an
excess of galaxies to the north-east with magnitudes $20<I<23$ and
colours $1.05<R-I<1.25$. This extension of red objects is also evident
in Figure~\ref{fig:grey} (which was compiled from independent data
from that used by Bremer et al.). The quasar has a dramatically
shorter southern lobe which is coincident with a spectrally confirmed
compact group of galaxies at the quasar's redshift.

\subsubsection{MRC~B1006--299, $z=1.064$}

The value of $B_{gq}$ for this field is large ($2092 \pm 1304$) and
there is a $4\sigma$ peak in the density of objects with $1.5 < R - z
< 2.0$ within 0.5 Mpc of the RLQ (Figure~\ref{fig:grey}).

\begin{figure}

  \begin{center}

    \epsfig{file=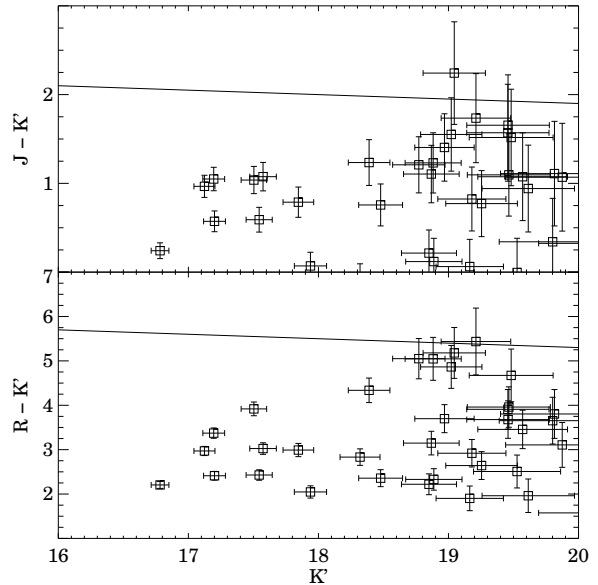,angle=0,width=8cm}

    \caption{The colour-magnitude diagram of objects in the
    \arcmn{2}{6} $\times$ \arcmn{2}{6} field around MRC~B0346--279. The thick
    line represents colour-magnitude relation for a cluster at $z
    =0.989$, based on the observed colours of Coma early-types
    (Stanford et al. 1998). No red sequence is seen.}

    \label{fig:0346_colmag}

  \end{center}

\end{figure}

\begin{figure}

  \begin{center}

    \epsfig{file=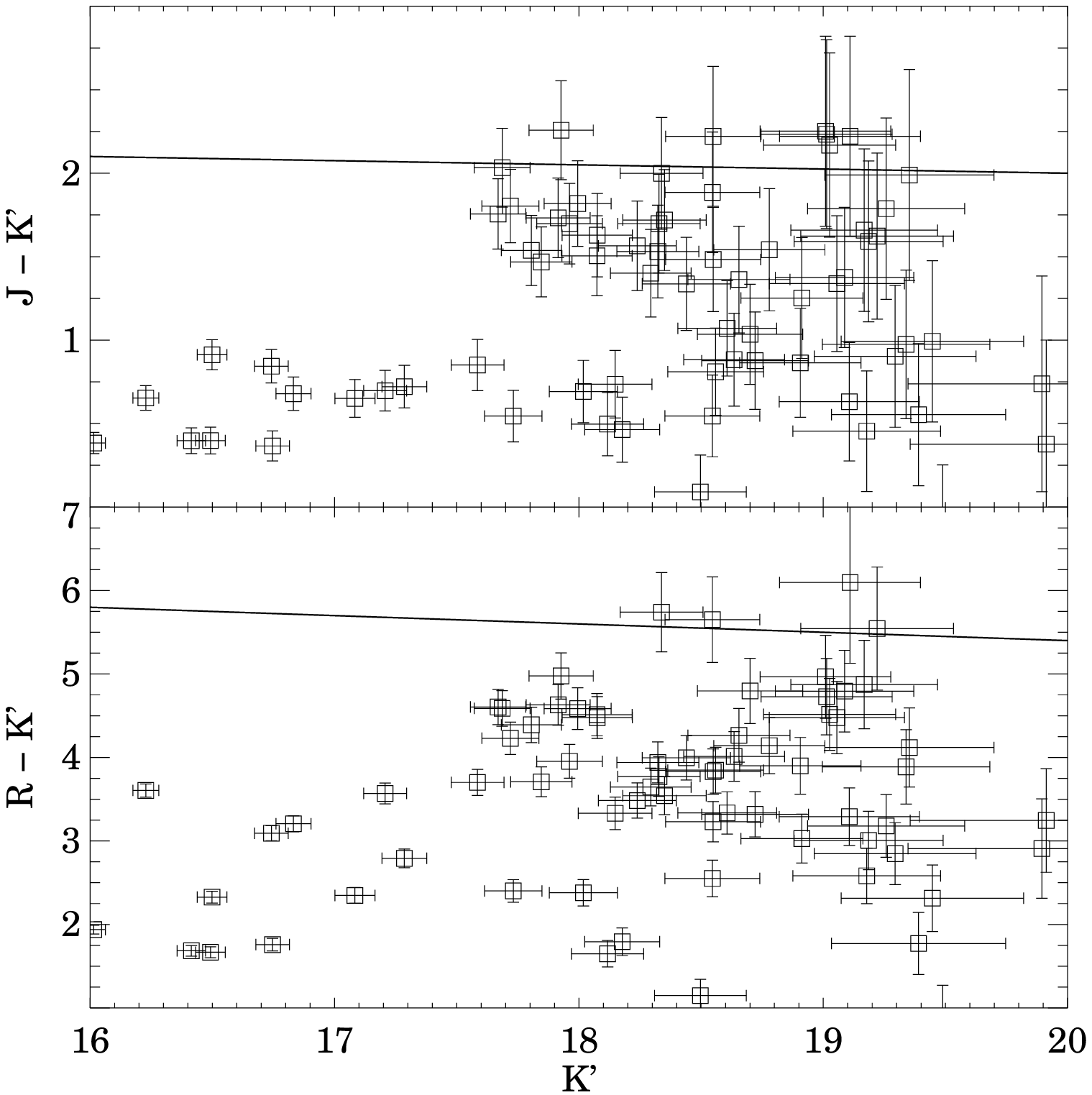,angle=0,width=8cm}

    \caption{The colour-magnitude diagram of objects in the
    \arcmn{2}{6} $\times$ \arcmn{2}{6} field around MRC~B1006--299. The
    thick line represents the no-evolution colour-magnitude relation
    for early-type galaxies at $z =1.064$, based on the observed
    colours of Coma early-types (Stanford et al. 1998).}

    \label{fig:1006_colmag}

  \end{center}

\end{figure}

\begin{figure}

  \begin{center}

    \epsfig{file=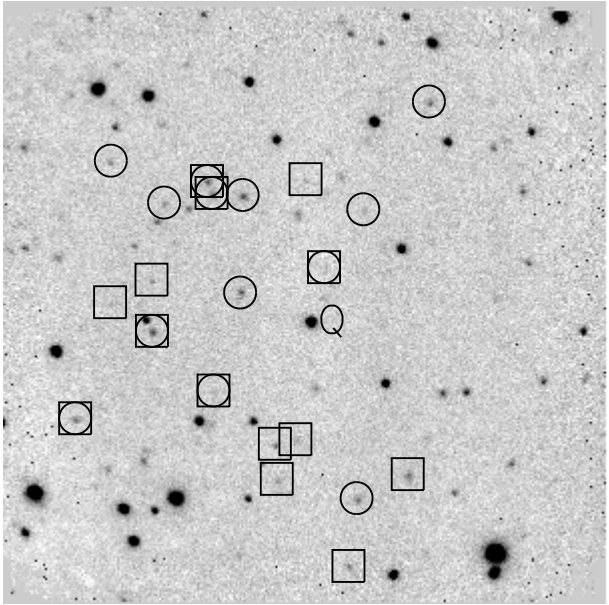,angle=0,width=8cm}

    \caption{Objects in the field of MRC~B1006--299 with
    $17<K^{\prime}<19$ and $4.3<R-K^{\prime}<4.9$ (circles),
    $1.7<J-K^{\prime}<2.3$ (squares). The image is the \arcmn{2}{6}
    $\times$ \arcmn{2}{6} $J$-band image and the quasar is labelled
    Q. North is up, East is left.}

    \label{fig:1006_ir}

  \end{center}

\end{figure}

The NIR evidence points to this overdensity being a lower redshift
system. Figure~\ref{fig:1006_colmag} shows a clear excess of objects
with $K^{\prime} \sim 18$ and $R - K^{\prime} \sim 4$, $J - K \sim
1.5$ compared with MRC~B0346--279
(Figure~\ref{fig:0346_colmag}). However, the $R-K^{\prime}$ {\em vs.}
$K^{\prime}$ red sequence expected for clusters at $z \sim 1$ is $\sim
1$ mag redder than these colours (shown as the solid line in
Figure~\ref{fig:1006_colmag}). Within the NIR field of view there are
13 objects with $4.3<R-K^{\prime}<4.9$, located mostly to the NE of
the quasar in the region of the $4\sigma$ optical overdensity (Figure
\ref{fig:1006_ir}). The magnitudes of these principal members are
$22.0<R<22.5$. These colours and magnitudes are indicative of bright
cluster members at $z \sim 0.8$ \cite{snellen96,aragon-salamanca98}.

\nocite{stanford98}

\nocite{stanford98}

There is clearly a system of associated galaxies in this field as
evinced by the proximity of objects both in space and colour. The
projected density of objects with $K>19$ and $3.0<I-K<3.8$ is 6 times
that found in the \arcmn{47}{2} Herschel deep field
\cite{mccracken00}.

The magnitude and optical-NIR colour of the clustering objects imply
that they are not at the same redshift as the quasar but at $z \sim
0.8$. Spectroscopy on 8m-class telescopes is required to fully
characterise the nature of this field. For the present, it is
classified as a quasar in a field environment with a cluster of
galaxies in the foreground.

\subsubsection{MRC~B1202--262, $z=0.786$}

The values of $B_{gq}$ and $N_{0.5}$ ($470 \pm 232$ and $4.2 \pm 5.1$)
are inconclusive. Figure~\ref{fig:grey} shows $3\sigma$ overdensities
of red galaxies NE and SW of the quasar. Unfortunately, without
further optical or NIR images of this field we are unable to determine
whether these overdensities are real systems of galaxies or chance
superpositions. Classification of the environment of MRC~B1202--262
will have to wait for deeper imaging or spectroscopy.

\subsubsection{MRC~B1208--277, $z=0.828$}

Figure~\ref{fig:grey} shows a $3\sigma$ overdensity of objects with
$V-I>2.0$ centred on the quasar.

\begin{figure}

  \begin{center}

    \epsfig{file=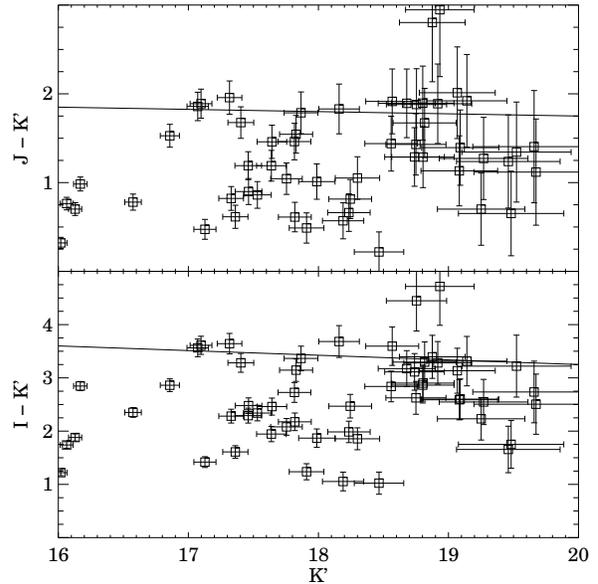,angle=0,width=8cm}

    \caption{The colour-magnitude diagram of objects in the
    \arcmn{2}{6} $\times$ \arcmn{2}{6} field around MRC~B1208--277. The
    thick line represents colour-magnitude relation for the cluster
    MS~1054.5-0321 at $z =0.828$, based on the observed colours of
    Coma early-types (Stanford et al. 1998).}

    \label{fig:1208_colmag}

  \end{center}

\end{figure}

There are $>4$ times as many galaxies with $K^{\prime}<19$ and
$3.0<I-K^{\prime}<3.8$ within the NIR field of view than are expected
from observations of the Herschel deep field
\cite{mccracken00}. Figure \ref{fig:1208_colmag} shows that these
objects make up a red-sequence consistent with a $z=0.828$ cluster of
galaxies.

\nocite{stanford98}

\begin{figure}

  \begin{center}

    \epsfig{file=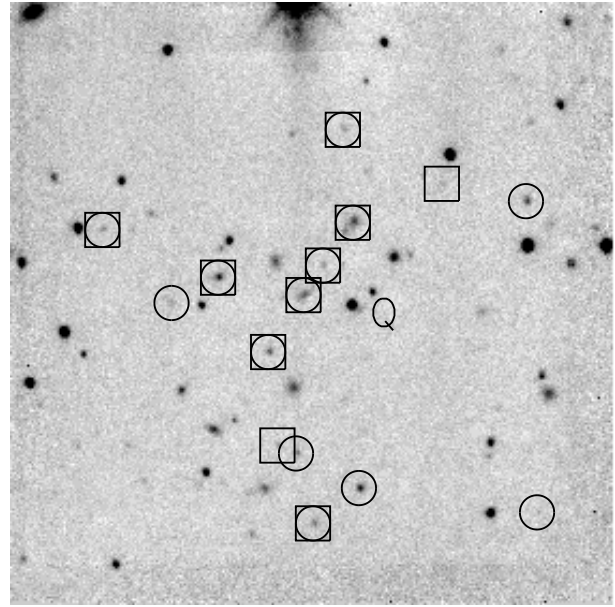,angle=0,width=8cm}

    \caption{Objects in the field of MRC~B1208--277 with
    $17<K^{\prime}<19$ and $3.0<I-K^{\prime}<3.8$ (circles),
    $1.7<J-K^{\prime}<2.1$ (squares). The image is the \arcmn{2}{6}
    $\times$ \arcmn{2}{6} $J$-band image and the quasar is labelled
    Q. North is up, East is left.}

    \label{fig:1208_ir}

  \end{center}

\end{figure}

\begin{figure}

  \begin{center}

    \epsfig{file=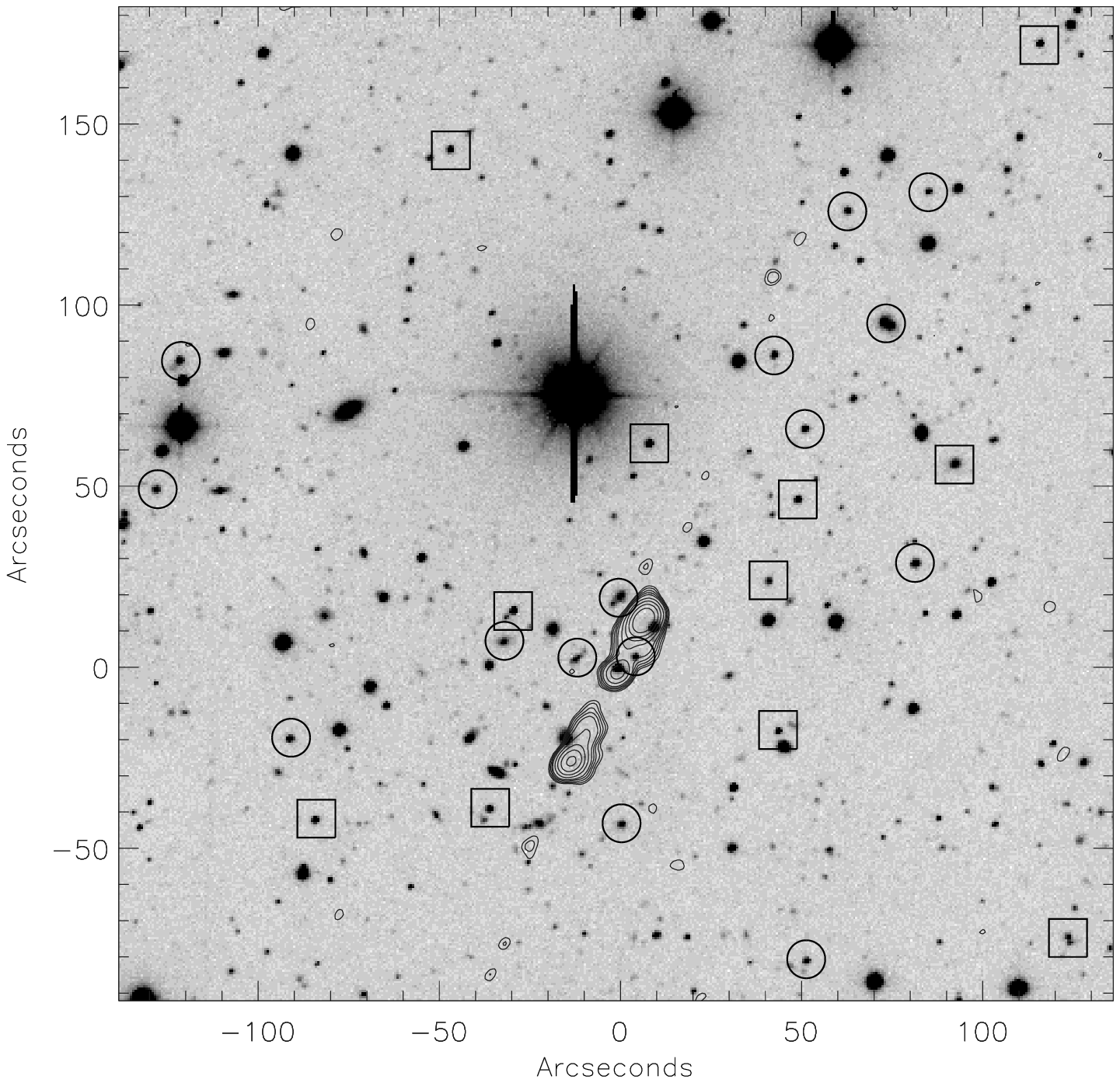,angle=0,width=8.5cm}

    \caption{The $I$-band image of the field of MRC~B1208--277
    overlaid with the 4860 MHz VLA radio contours. Indicated are
    objects with $19<I<21$ and $2.0<V-I<2.5$ (circles) and
    $2.5<V-I<3.0$ (squares). North is up, East is left and the quasar
    is at 0,0. The large-scale structure of galaxies appears aligned
    with the radio lobes in a SE-NW direction.}

    \label{fig:1208_align}

  \end{center}

\end{figure}

Figure~\ref{fig:1208_ir} and Figure~\ref{fig:1208_align} show that
candidate cluster members display a tendency to align themselves NW --
SE. This is the same direction as the radio lobes (shown in
Figure~\ref{fig:1208_align}). Five of the objects so distributed are
close doubles or have a disturbed morphology.

The $V$-band magnitudes of candidate cluster members display a greater
variation than expected of a cluster of galaxies at this
redshift. Observations of MS~1054--03 indicate that core ellipticals
have $V-I \sim 2.9$ \cite{vandokkum01} at $z=0.828$. Candidate core
cluster galaxies around MRC~B1208--277 have median $V-I \sim 2.4$, \ie
half a magnitude bluer than ellipticals in MS~1054--03. Small amounts
of on-going star formation can affect the $V$-band magnitude at these
redshifts and could explain this discrepancy.

\subsubsection{MRC~B1224--262, $z=0.768$}

A $5\sigma$ overdensity is seen in Figure~\ref{fig:grey} with
$2.5<V-I<3.0$, the colour of passively-evolving ellipticals at
$z=0.768$.

\begin{figure}

  \begin{center}

    \epsfig{file=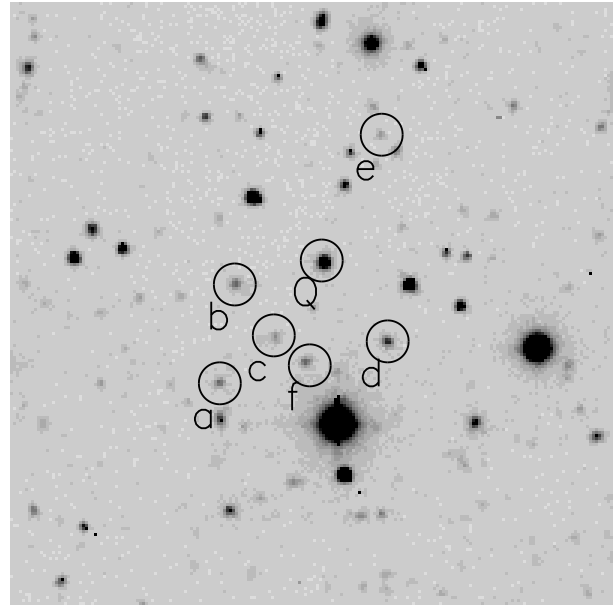,angle=0,width=8cm}

    \caption{Co-added $V,R,I$ image of the central $2^{\prime} \times
    2^{\prime}$ around MRC~1224--262 (labelled Q). Objects with
    $20.5<I<22.5$ and $4.5<R-K^{\prime}<5.2$ are indicated. The area
    shown here is that covered by the NIR field of view, North is up,
    East is left.}

    \label{fig:1224_3}

  \end{center}

\end{figure}

There are 6 objects within 0.5 Mpc with $20.5<I<22.5$ and
$4.5<R-K^{\prime}<5.2$ and $1.38<R-I<1.58$ (Figure
\ref{fig:1224_3}). The tightness of these colours, as well as the
colours themselves point to the fact that these galaxies are located
in the same system at the redshift of the quasar.

\begin{figure}

  \begin{center}

    \epsfig{file=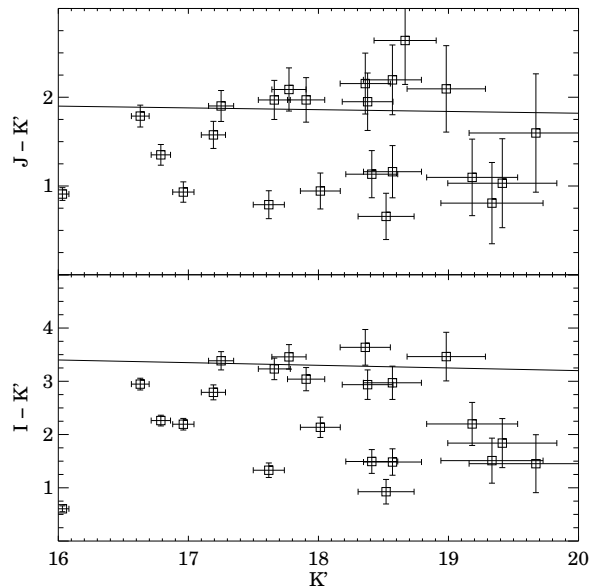,angle=0,width=8cm}

    \caption{The colour-magnitude diagram of objects in the
    \arcmn{2}{6} $\times$ \arcmn{2}{6} field around MRC~B1224--262. The
    thick line represents no-evolution colour-magnitude relation for
    early-type galaxies at $z =0.768$, based on the observed colours
    of Coma early-types (Stanford et al. 1998).}

    \label{fig:1224_colmag}

  \end{center}

\end{figure}

This is a sparse field, as evinced by the total number of objects in
the colour-magnitude diagram. Nevertheless, it is possible to identify
a red sequence in $I-K^{\prime}$ and $J-K^{\prime}$ in
Figure~\ref{fig:1224_colmag}.

\nocite{stanford98}

The number of objects within the central square arcminute around the
AGN with $2.6<I-K<3.8$ and $K<19$ is seven times that expected from
analysis of the Herschel deep field \cite{mccracken00}. It must also
be borne in mind that our data is incomplete above $K^{\prime} = 18.5$
and so fainter objects in the field are missed; thus the true
overdensity is likely to be greater than this.

MRC~B1224--262 resides in an overdensity of galaxies consistent with
being at $z=0.768$. We have identified candidate core cluster members,
(Figure~\ref{fig:1224_3}, and a possible red sequence,
Figure~\ref{fig:1224_colmag}). Although there are relatively few of
these, this system does not satisfy the modified Hickson isolation
requirement and thus is not classified as a compact group. The
overdensity around MRC~B1224--262 may be a loose grouping or a poor
cluster of galaxies.

\subsubsection{MRC~B1301--251, $z=0.952$}

\begin{figure}

  \begin{center}

    \epsfig{file=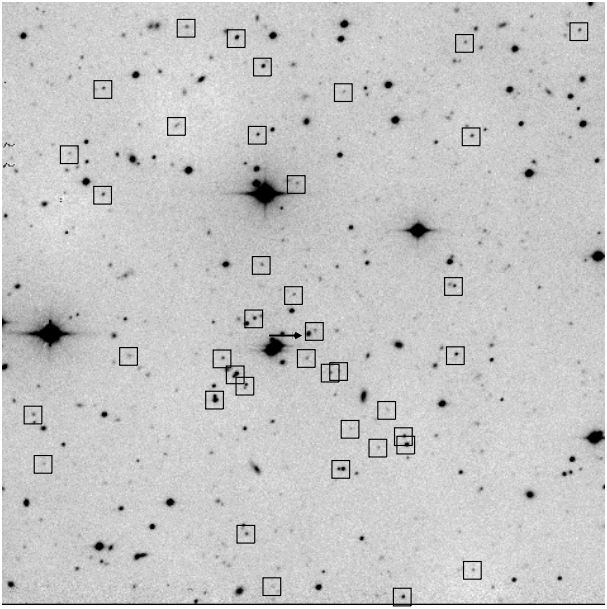,angle=0,width=8cm}

    \caption{Objects in the field of MRC~B1301--251 with
    $19 < z < 23$ and $2.8 < V - z < 3.4$. The quasar is
    marked with the horizontal arrow. The field of view is
    \arcmn{4}{9} $\times$ \arcmn{4}{9}, North is up, East is left.}
 
    \label{fig:1301_objs}

  \end{center}

\end{figure}

No estimate of $B_{gq}$ is possible because of the insufficient depth
of the $I$-band image. However, a $4\sigma$ overdensity in
Figure~\ref{fig:grey} and the spatial distribution of galaxies with
the colours of passive ellipticals at $z=0.952$ in Figure
\ref{fig:1301_objs} indicate clustering of galaxies centred on the
quasar.

\begin{figure}

  \begin{center}

    \epsfig{file=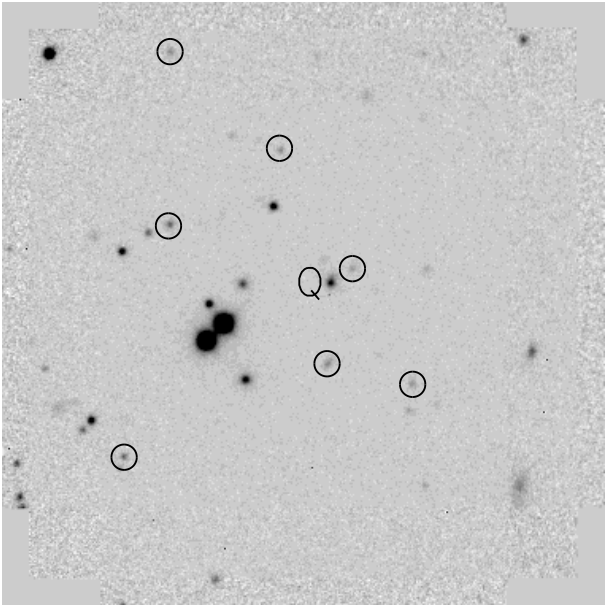,angle=0,width=8cm}

    \caption{The $H$-band image of the central arcminute about
    MRC~B1301--251, North is up, East is left. Indicated are those
    objects with $1.8<I-H<3.0$ and $H<21$. The quasar is marked with
    the Q.}

    \label{fig:1301_ir}

  \end{center}

\end{figure}

The $H$-band image only covers an area of $1$ square arcminute, as
shown in Figure \ref{fig:1301_ir}. Nevertheless, there are 7 objects
with $19<H<21$ and $1.8<I-H<3.0$. This density is three times that of
objects with similar colours in the Herschel deep field (assuming
$H-K^{\prime} = 1$ for ellipticals at $z=0.952$). The number and
density of these objects around MRC~B1301--251 point to the presence
of a cluster of galaxies.

\subsubsection{MRC~B1303--250, $z=0.738$}

Figure~\ref{fig:grey} shows a $3\sigma$ peak in the density contrast
to the east of the RLQ. VLA observations show a bent radio source
\cite{kapahi98}, possibly indicating interaction with an ICM. However,
as is the case with MRC~B1202--262, this is the extent of the
information on this source. MRC~B1303--250 may reside on the edge of a
cluster of galaxies, however more detailed imaging/spectroscopy will
be required before this can be confirmed.

\subsubsection{MRC~B1355--236, $z=0.832$}

\begin{figure}

  \begin{center}

    \epsfig{file=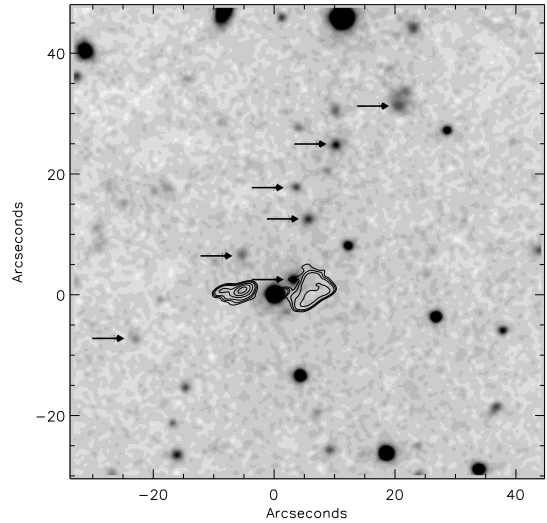,angle=0,width=8.4cm}

    \caption{The central $\sim 1^{\prime}$ of the $I$-band image of
    MRC~B1355--236 overlaid with the 4860 MHz VLA radio
    contours. Objects with $20<I<23$ and $1.2<R-I<1.5$ are
    indicated. North is up, East is left, the quasar is at 0,0.}

    \label{fig:1355_rad}

  \end{center}

\end{figure}

There is a linear clustering of objects to the NW of the
quasar. Figure \ref{fig:1355_rad} shows seven objects with $20<I<23$
and $1.2<R-I<1.5$, consistent with being $z=0.832$ ellipticals. The
most north-western member of the system has a very diffuse morphology
and may be a merger. There is marginal ($3\sigma$) clustering of
objects with these magnitudes and colours to the East of the quasar
(Figure~\ref{fig:grey}). For this reason the system does not conform
to Hickson's isolation criterion and so could be a cluster core or a
compact group on the edge of a poor cluster rather than an isolated
compact group. Wide-field NIR imaging or MOS is required to classify
the environment of MRC~B1355--236 more accurately.

\subsection{Summary}

The fields of all 21 quasars in this sample were examined for signs of
clustering. Objects with $>3\sigma$ overdensities in the density maps
were set aside as cluster candidates and examined further. Those
fields with $>5\sigma$ overdensities, or $>3\sigma$ with an
enhancement of NIR number counts over field estimates were classified
as clusters of galaxies. The fields were then examined to see if they
complied with the modified Hickson requirements to qualify as a
compact group.

The analysis reveals clustering of galaxies in the fields of 8
quasars. In two cases, MRC~B0058--229 and MRC~B0941--200, the
clustering is consistent with being a compact group. Poor or missing
data meant that no determination can be made as to the nature of 3
fields (labelled `?' in Table~\ref{tab:bgq}). In two cases
(MRC~B0346--279 and MRC~B1006--299) high values of $B_{gq}$ and
$N_{0.5}$ are likely to be caused by intervening systems of
objects. All other (8) objects have environments consistent with
other galaxies in the field. The classifications are listed in
Table~\ref{tab:bgq}.

The mean $B_{gq}$ values of the clustered systems (C/G) are
systematically richer than those of the fields (F), $436 \pm 178$ as
opposed to $159 \pm 150$. However, there is considerable overlap in
the statistics when individual fields are examined. This suggests that
$B_{gq}$ is only useful as an ensemble statistic and is not
necessarily an accurate predictor of a single high-redshift quasar's
environment. For a further discussion of this point see
\S\ref{sec:sing_crap}.

\section{Discussion}
\label{sec:disc}

\subsection{Clustering in the fields of RLQs}

We find strong evidence for clustering of galaxies about quasars in
the fields of 8 of the 21 RLQs (C or G in Table~\ref{tab:bgq}). The
nature of the clustering varies from compact groups to rich
clusters. Marginal evidence for clustering is seen in 3 other fields
(labelled as `?' in Table~\ref{tab:bgq}). In two further cases,
MRC~B0346--279 and MRC~B1006--299, suspected intervening systems of
galaxies are found. The rest of the quasars have environments
consistent with field galaxies.

The clusters of galaxies found in this work are selected based on
their old elliptical population. This means that in the systems at
$0.6 < z < 1.1$ detailed in \S\ref{sec:indi}, the passively-evolving
elliptical core galaxies are already a significant constituent. Other
studies of (richer) clusters of galaxies at these redshifts (\eg
Stanford \etal 1998; van Dokkum \& Franx
2001)\nocite{stanford98,vandokkum01} find a similar state of
affairs. Results such as these have been used to argue that the
formation epoch of the majority of stars in early-type galaxies is $z
\sim 2 - 3$ (van Dokkum \& Franx 2001\nocite{vandokkum01}; see also
\S\ref{sec:ccg}).

Targetting quasars is a very efficient way of finding structure at
high redshift.  We find compact groups or clusters of galaxies in at
least $\sim 40\%$ and possibly $> 50\%$ of our RLQ fields, a return of
$\sim 40$ clustered systems per square degree imaged ({\em nb:} not
$\sim 40$ clusters per square degree of sky). This compares favourably
with current high-redshift cluster surveys, \eg $\sim 1 - 2$ clusters
at $z \lesssim 1.2$ per square degree for the Red-Sequence Cluster
Survey \cite{gladders00} or the predicted $\sim 10$ clusters per
square degree at $0.6 < z < 1.1$ for the XMM-LSS
\cite{romer01,refregier02}.

We have shown that detecting clustered systems above a redshift of 0.6
with 4m data and single-filter techniques is difficult. With the
addition of colour information and NIR filters it is possible to
extend the searches above $z = 1$. These limits are comparable in
terms of depth with the most recent wide-field optical and X-ray
surveys. In order to study clustered systems at redshifts of 1 to 2 or
greater, 8m data will be required. Observations in this redshift range
are desirable because it represents the peak in the star formation
activity of the universe (Madau, Pozzetti \& Dickinson
1998)\nocite{madau98}, the period of maximum AGN activity
\cite{hartwick90,peterson97} and possibly the epoch of major cluster
assembly. It is likely that the interplay between these distinct
phenomena can be disentangled using observations of clustered fields
at $z \sim 2$. However, most 8m telescopes have cameras with a field
of view of $\sim 5^{\prime}$. It would therefore take a prohibitively
long time to complete a survey with these facilities. The XMM-LSS will
detect systems at $z = 1 - 2$ but numbers will be weighted
significantly toward $z = 1$, with few at $z \sim 2$. Until
coordinated X-ray and SZ searches come on line, the only way to detect
systems of galaxies efficiently at redshift $\sim 2$ is to use
targetted searches.

Imaging the fields of quasars also represents a means of compiling
complimentary catalogues to those of other cluster surveys. The fields
of RLQs are typically poorer than the systems identified by optical or
X-ray means, and they are likely to be less massive than the clusters
SZ observations will discover. However, in common with systems
detected by these means, our clusters display no evidence for
evolution to $z \gtrsim 0.9$ (Lubin 1996; Ebeling \etal 1998, but see
Gioia \etal 2001) \nocite{lubin96,ebeling98,gioia01}. We also detect
the presence of a red sequence in our clusters. This indicates that
like richer clusters, their passively-evolving ellipticals formed at
$z \gtrsim 2$ \cite{stanford98,holden01,vandokkum01}.

\subsection{Colours of core cluster galaxies}
\label{sec:ccg}

\begin{figure*}

  \begin{center}

    \hbox{
    \epsfig{file=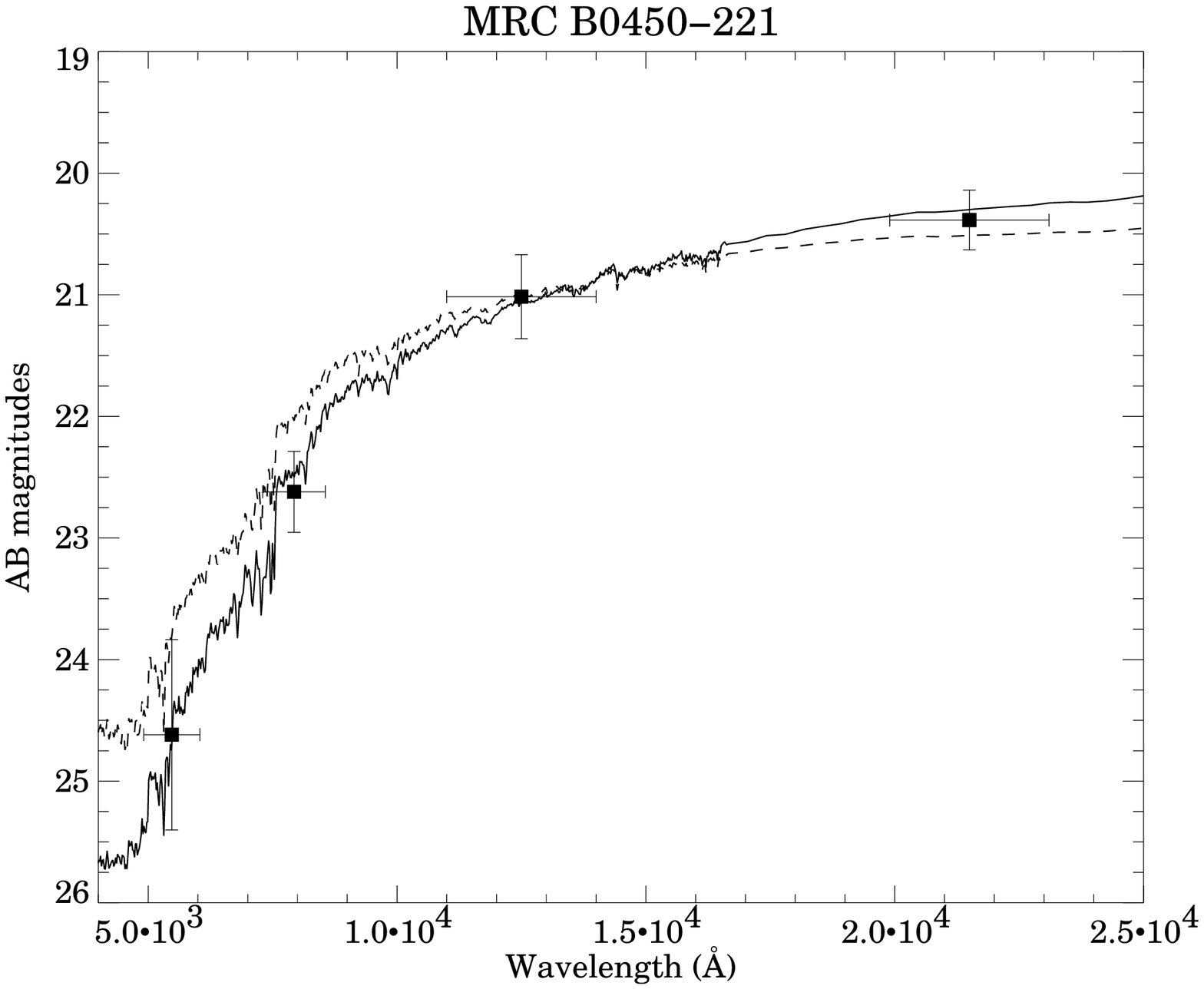,width=8cm,angle=0}
    \epsfig{file=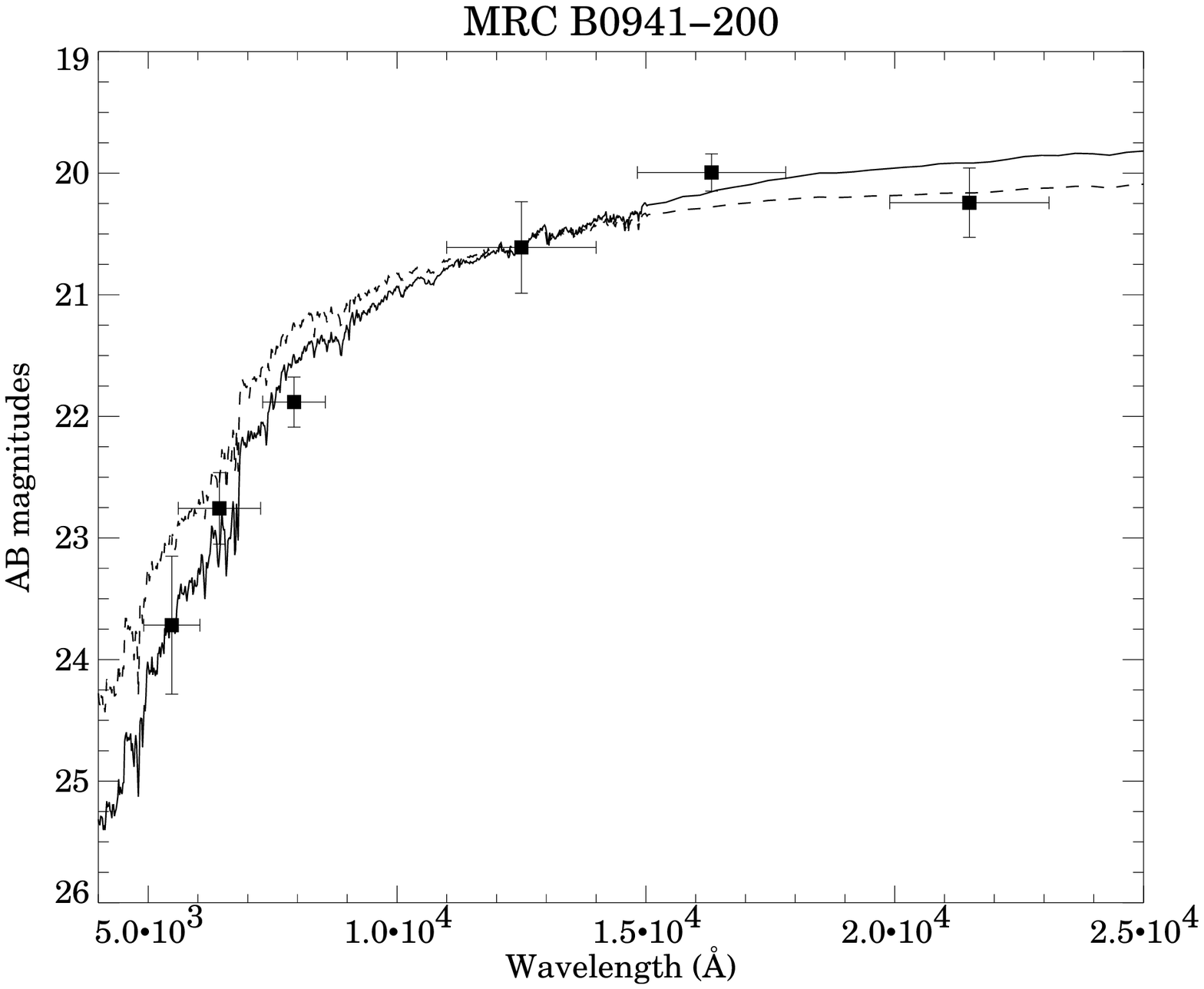,width=8cm,angle=0}
    }

    \hbox{
    \epsfig{file=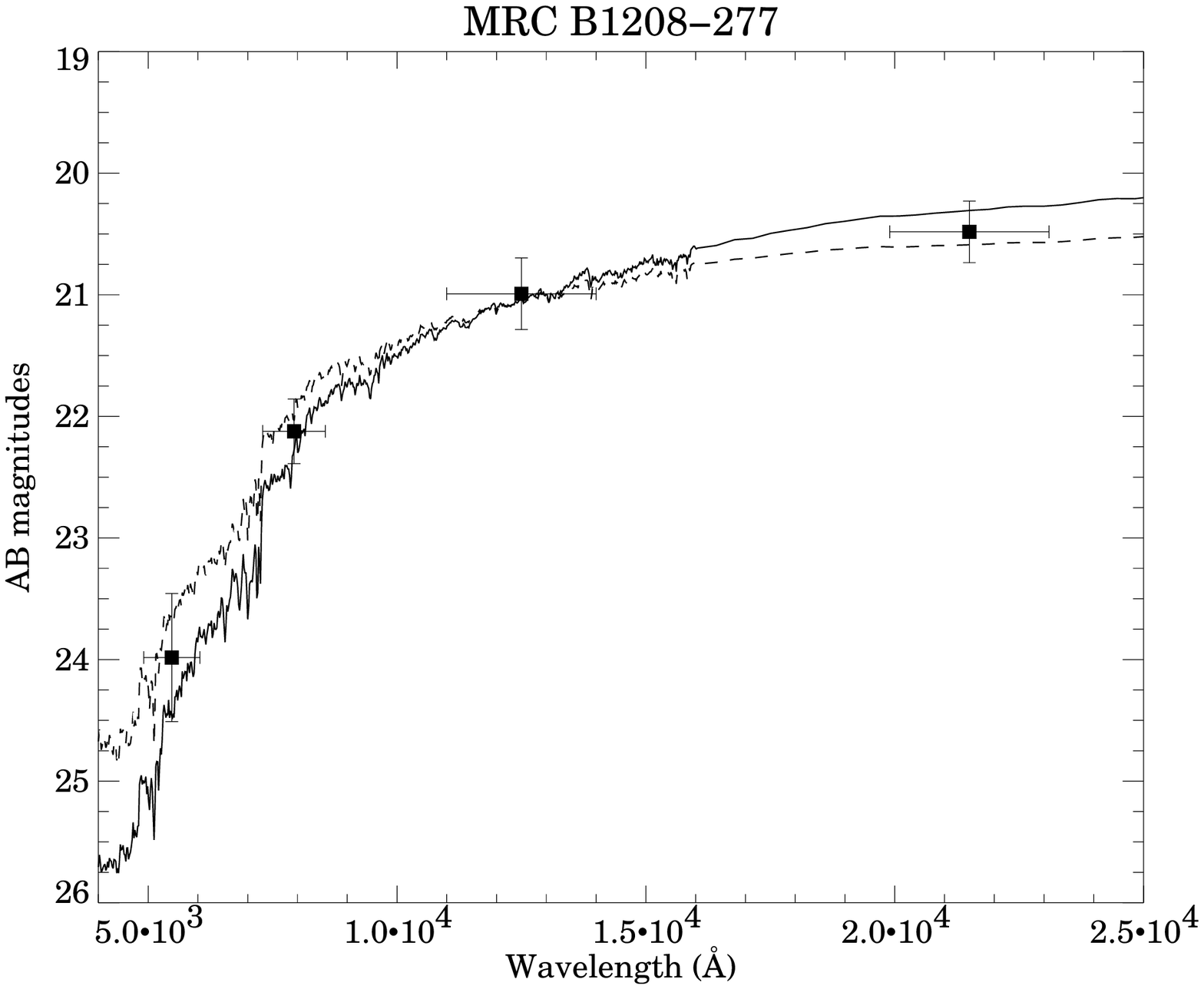,width=8cm,angle=0}
    \epsfig{file=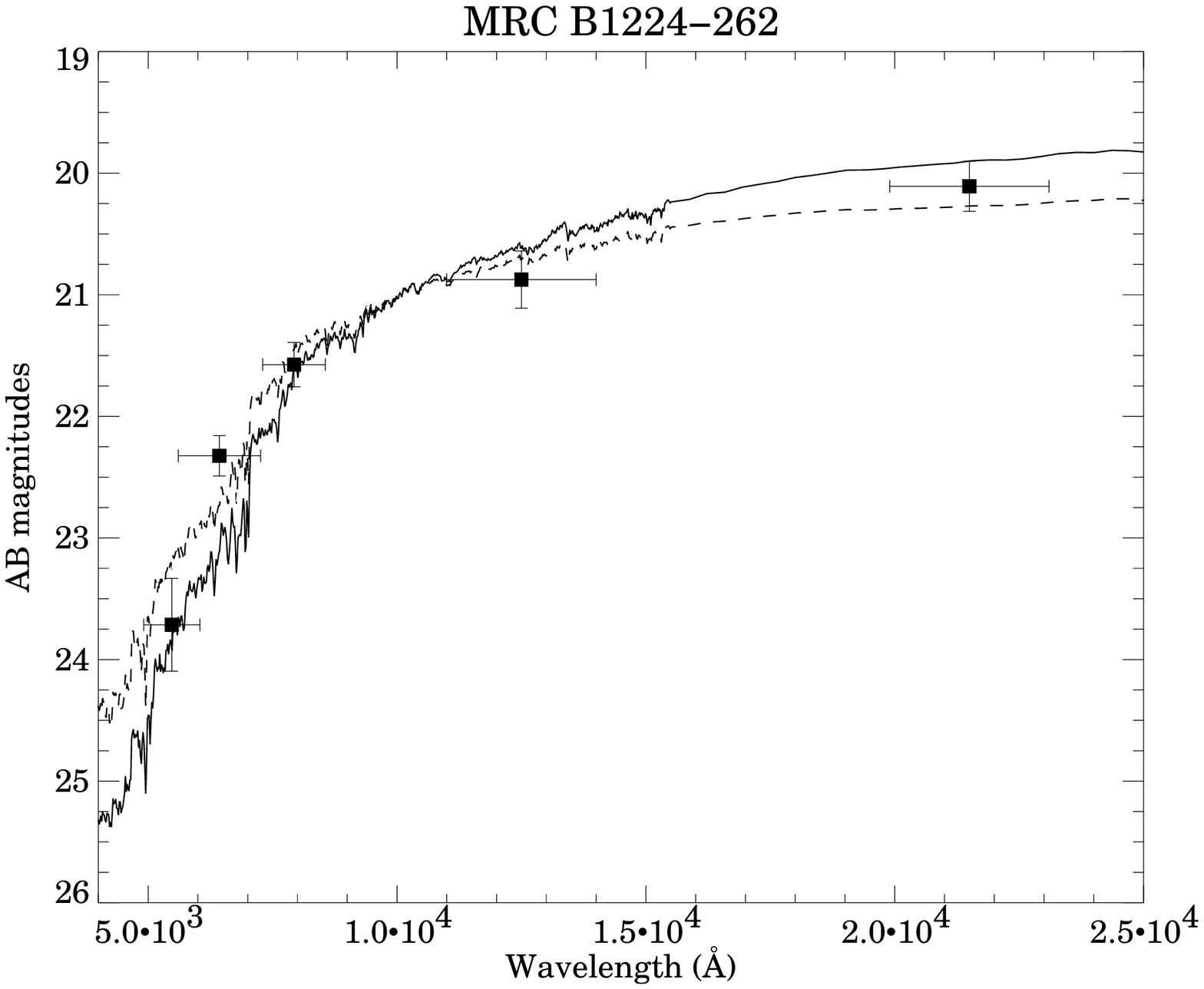,width=8cm,angle=0}
    }

    \caption{Average SEDs for red objects in the fields of
    clusters. The data points are compiled from the objects with the
    colours of passively-evolving ellipticals at the quasar redshift
    (see Figure~\ref{fig:grey} for the precise selection criteria)
    calibrated to AB magnitudes. Model elliptical galaxy spectra from
    Fioc \& Rocca-Volmerange (1997) are shown with a $\lambda^{-1.3}$
    reddening dependence added. The solid line represents a 6 Gyr, and
    the dashed line a 2 Gyr old galaxy.}

    \label{fig:sed}

  \end{center}

\end{figure*}\nocite{fioc97}

Assuming that core cluster galaxies have certain colours carries a
central disadvantage, namely that minimal analysis can be undertaken
to investigate the properties of these colours. However, some analyses
can be made. One important test involves the spectral energy
distributions of galaxies and is featured in Figure~\ref{fig:sed}.

There are four cluster fields for which at least two optical and at
least two NIR images exist. It is these latter NIR filters that are
essential to provide enough of a baseline over which to test different
SEDs. Data points in each filter were assembled by considering only
objects with the colours of passively-evolving ellipticals at the
quasar redshift, \ie those galaxies used to construct the panels in
Figure~\ref{fig:grey}. These were then converted to AB magnitudes
using the prescriptions of Fukugita \etal (1995)\nocite{fukugita95}.

The spectra were assembled from the models of Fioc \& Rocca-Volmerange
(1997)\nocite{fioc97}, redshifted to the appropriate epoch and scaled
to fit the points as well as possible. Six Gyr and two Gyr elliptical
models were considered, both with solar metallicity and an
exponentially decreasing star formation rate. There are no good fits
to the resulting spectrum unless some host galaxy reddening is
applied. Bremer \etal (2002)\nocite{bremer02} found that spectroscopic
observations of the galaxies in the MRC~B0941--200 group can be made
to agree well with a 3 Gyr model if a $\lambda^{-1.3}$ reddening
dependence is introduced. This is therefore applied to the models,
normalising to $A_{V} = 0.4$ for consistency with observations.

The resulting SED comparisons are shown in Figure~\ref{fig:sed}. The
data is well fit by elliptical galaxies with stellar masses of $\sim
10^{11}$\msun, ages of $2 - 6$ Gyr and on-going star formation rates of
$0.5 - 2$ \msun yr$^{-1}$. For the fields of MRC~B0450--221 and
MRC~B0941--200 the data points favour the older model while in the
other two fields no preference is found.

These data indicate that at $0.7<z<0.9$, the elliptical population is
at least $2$ Gyr old and potentially $6$ Gyr old. This suggests that
the passively-evolving galaxies in the cores of the clusters form at
redshifts of $2 - 3$. If passively-evolving ellipticals form at $z
\gtrsim 2 $ and are in place at $z \sim 1$ the indications are that a
cluster will not have had time to virialise. The time between these
redshifts is $\sim 3 - 4$ Gyr whereas virialisation timescales are
typically many Gyr for cluster mass scales. Even at $z = 0$, the outer
parts of many clusters are not dynamically relaxed.  Certainly, the
distributions of the ellipticals about the RLQs and the positions of
the quasars in their putative clusters (see
\S\ref{sec:g_d}) suggest that these systems are not virialised.

\subsection{The distribution of galaxies around RLQs}
\label{sec:g_d}

\begin{figure}

  \begin{center}

    \epsfig{file=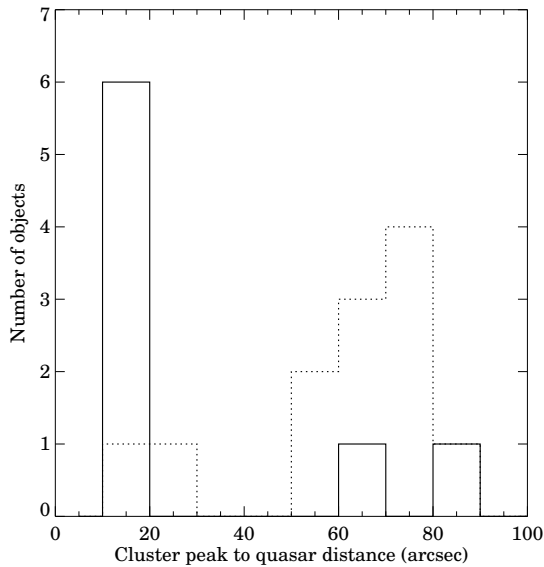,angle=0,width=8cm}

    \caption[Histogram of the projected distance between cluster peaks
    and RLQs]{{\em Thick line:} Histogram showing the projected
    distance between the peak value of the clustering statistic (as
    calculated in \S\ref{sec:mult}) and the RLQs denoted `C' or
    'G'. {\em Dotted line:} The same but for those quasars categorised
    as `F' (Table~\ref{tab:bgq}). In both cases the distance is
    calculated only within 500kpc of each quasar.}

    \label{fig:c_dis}. 

  \end{center}

\end{figure}

In nearly all cases in which clustering is evident, there is a $\sim
10^{\prime\prime} - 20^{\prime\prime}$ distance between the RLQ and
the centroid of the red galaxy distribution. This is illustrated
graphically in Figure~\ref{fig:c_dis}. The figure shows the projected
distance between the peak value of the clustering statistic (as
calculated in
\S\ref{sec:mult}) and the RLQs. These distances are only determined
for points within 500kpc of each quasar. If there is no overall
correlation between the quasar position and cluster peak, the
peak-to-RLQ distance should increase steadily to a maximum value of
$\sim 60^{\prime\prime} - 80^{\prime\prime}$, a reflection of the
greater area sampled at increased radii. This is exactly what happens
to the quasars in field environments. However, RLQs marked `C' or `G'
in Table~\ref{tab:bgq} show a marked tendency to exist
$10^{\prime\prime} - 20^{\prime\prime}$ away from the cluster peak. Of
the two quasars which do not obey this relationship, there is evidence
that both MRC~B0941--200 and MRC~B1355--281 exist in a compact group
at the edge of a larger-scale structure (\S\ref{sec:indi}).

\begin{figure}

  \begin{center}

    \epsfig{file=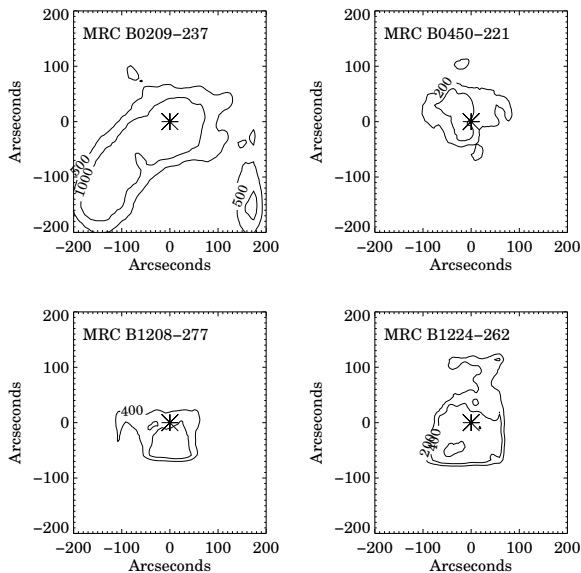,angle=0,width=8cm}

    \caption[The spatial cross-correlation amplitude in the fields of
    four quasars which reside in clusters]{The spatial
    cross-correlation amplitude at all points in the fields of four
    quasars which reside in clusters. The quasars are marked by an
    asterisk, North is up, East is left.}

    \label{fig:bgq_map}

  \end{center}

\end{figure}

The panels in Figure~\ref{fig:grey} and Figure~\ref{fig:bgq_map} also
indicate that quasars in the sample are not necessarily found at the
centres of galaxy clusters. However, without X-ray data, it is not
possible to tell whether an AGN truly resides at the edge of a cluster
or whether the galaxy distribution fails to follow the gravitational
potential faithfully. The optical study of Cygnus A and its
surrounding cluster of galaxies by Owen \etal (1997)\nocite{owen97}
shows that the powerful radio source is offset by $\sim
10^{\prime\prime}$ from the galaxy distribution. However, in this case
the radio galaxy is found at the bottom of the potential well, as
traced by the X-ray emitting gas \cite{smith02}.

Clustering of galaxies also extends further than $100^{\prime\prime}$
($\sim 1$ Mpc) from the RLQs, \eg MRC~B0030--220, MRC~B0058--229,
MRC~B0941--200, MRC~B1208--277 in Figure~\ref{fig:grey}. Similar
results have been observed by S\'anchez and Gonz\'alez-Serrano (1999;
2002)\nocite{sanchez99,sanchez02} for (non colour-selected) galaxies
in the fields of 7 RLQs at $1.0<z<1.6$.

In general, the distribution of passive ellipticals in the clustered
systems is irregular. None of the agglomerations display obvious
spherical symmetry. This may be a product of shallow sampling, or may
reflect the genuine topology of the systems. The fact that the quasars
themselves (and by implication their massive elliptical hosts) are not
centred on the elliptical distribution may indicate the second of
these options. Other observations have suggested that {\em rich}
high-redshift clusters can display an irregular X-ray morphology
\cite{donahue98}, indicating further that the poorer clustered
systems in this work are likely to be unvirialised, or are made up of
several smaller-scale virialised components.

The fact that RLQs are not found at the centre of galaxy clusters and
the projected extension of these distributions can be used to explain
the results of Hall \& Green (1998)\nocite{hall98b}. Hall \& Green
found that overdensities of $K \gtrsim 19$ around $1 < z < 2$ RLQs
occurred on two spatial scales. The first was a significant
enhancement at $< 40^{\prime\prime}$ from the quasars compared with
the rest of the field. The second was a higher density over the $\sim
100^{\prime\prime}$ field of view compared with literature values.

\begin{figure}

  \begin{center}

    \epsfig{file=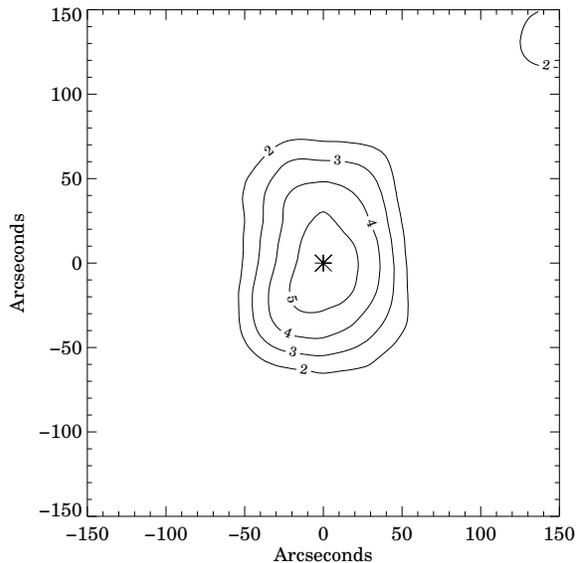,angle=0,width=8cm}

    \caption{Contour plot showing the density of objects in multiple,
	simulated clusters. Each cluster is created by placing objects
	randomly in a $50^{\prime\prime}$ radius and offset by up to
	$\sim 50^{\prime\prime}$ from the quasar. The results are
	stacked and the average density is found as with
	Figure~\ref{fig:grey} (see text). The contours are in units of
	$\sigma$.}

    \label{fig:c_sim}

  \end{center}

\end{figure}

Asymmetric clustering of galaxies about RLQs can, when the results are
combined, reproduce this sort of distribution. Figure~\ref{fig:c_sim}
shows the result of simulated clusters placed in the fields of
quasars. In 50 out of 100 cases, `clusters' of $\sim 20$ objects were
placed with their centre anywhere within $50^{\prime\prime}$ of a
`quasar'. This quasar was itself the centre of a `field' of about 20
objects randomly placed in a $300^{\prime\prime}$ square area. The
numbers were chosen to reproduce roughly the $K$ band number counts
which were the basis of Hall \& Green's selection. The resulting
distributions of objects were stacked and the density of objects was
calculated in the same manner as
Figure~\ref{fig:grey}. Figure~\ref{fig:c_sim} shows a $>4\sigma$
overdensity within $40^{\prime\prime}$ of the quasar which resembles
Hall \& Green's $4.5\sigma$ overdensity within $40^{\prime\prime}$
rather closely. An overdensity of objects over the whole field
relative to the background counts is also expected. The edge of
particular cluster can reach to $\sim 100^{\prime\prime}$ from the
quasar thereby producing an overdensity, albeit of low significance,
relative to the field.

The simulations suggest that Hall \& Green's results can be reproduced
if, as found with the RLQs in this work, about half the fields have
galaxy excesses asymmetrically distributed within $\sim
50^{\prime\prime}$ of the quasar.

\subsection{Single-filter clustering statistics}
\label{sec:sing_crap}

We have calculated $B_{gq}$ and $N_{0.5}$ for our sample of objects
and find them to be consistent with previous, lower redshift
studies. We find no evidence for evolution in clustering, measured
thus, with redshift. The assertion that there is large variance in the
clustering of galaxies around quasars is also supported by these
results.

Five of the eight groups and clusters of galaxies found using
multicolour means (labelled `C' or `G' in Table~\ref{tab:bgq}) have
insignificant or incalculable values of $B_{gq}$. This partly reflects
the fact that these systems are poorer. However, it is also due to the
lessening of the efficiency of the statistic with epoch. There are
various reasons why this occurs. At the highest redshifts ($z>0.9$)
the depth of the data only allows imaging of the brightest cluster
members. These several galaxies then do not constitute a significant
population and may be overwhelmed by foreground galaxies at the same
magnitude levels.

In attempting to compare clusters over a range of $z$, investigators
assume that the cluster morphology is independent of redshift. In
particular, when interpreting $B_{gq}$ and $N_{0.5}$ as a measure of
cluster richness, it is assumed that the quasar resides at the centre
of the galaxy distribution. It is known, however, from studies of the
most powerful high-redshift X-ray clusters that these systems may be
less regular than their low-redshift counterparts and exhibit
considerable sub-structure (\eg MS~1054-03; Donahue \etal
1998\nocite{donahue98}; van Dokkum \& Franx
2001\nocite{vandokkum01}). Section~\ref{sec:g_d} also showed that the
quasar does not reside in the centre of a galactic overdensity. This
means that $B_{gq}$ or $N_{0.5}$ can always be made greater by placing
the centre of the 0.5 Mpc circle used to calculate $A_{gq}$ somewhere
other than on the quasar. Figure \ref{fig:bgq_map} illustrates this
graphically. The spatial cross-correlation function is calculated for
four clusters in the sample. $B_{gq}$ is calculated at all points
within a 200$^{\prime\prime}$ square centred on the quasar assuming
the clustering is at the redshift of the quasar. In all cases in
Figure \ref{fig:bgq_map} it is evident that the value of the spatial
cross-correlation amplitude is not maximal at the quasar position but
is always offset by some 20$^{\prime\prime}$ to 50$^{\prime\prime}$
({\em cf:} also Figure~\ref{fig:c_dis}). This will lead to a
systematic underestimate in the richness of a system. At lower
redshift it has been assumed that the quasar is found in the central
cluster galaxy and thus that an Abell richness can be derived from
$B_{gq}$. This work shows that this is not the case and at $z>0.6$ it
seems likely that an Abell class cannot be reliably be assigned to a
particular value of $B_{gq}$.

Intervening systems of galaxies also distort the values of $B_{gq}$
and $N_{0.5}$. $B_{gq}$ for 2 of the 21 quasars in the sample,
MRC~B0346--279 and MRC~B1006--299, would lead them to be classified as
rich clusters at redshifts 0.989 and 1.064
respectively. Section~\ref{sec:indi} has shown that these
overdensities are not consistent with clusters of galaxies at these
redshifts. This suggests that $\sim 10\%$ of values of $B_{gq}$ for
quasars at $z \sim 1$ are overestimated due to the presence of
foreground systems.

Selecting clusters of galaxies by the red colours of their passive
elliptical population circumvents some of the problems mentioned
above. The colours of these galaxies are rare (or at least are made so
by the appropriate choice of filters), so dilution by faint foreground
galaxies becomes much less of an issue. The multicolour method also
has the power to reject intervening systems.

\subsection{Radio Properties}

A full discussion of the radio properties of the sample will be
presented in Barr \etal (2003, in preparation). For the time being, we
note two clustering trends involving the radio morphology of our
sample. Firstly that clustered environments are preferably associated
with sources whose sizes are $<200$kpc, while larger radio sources
are found in unclustered environments. Secondly, asymmetric radio
sources are almost always associated with clustered environments,
though not necessarily rich clusters.

\begin{figure}

  \begin{center}

    \epsfig{file=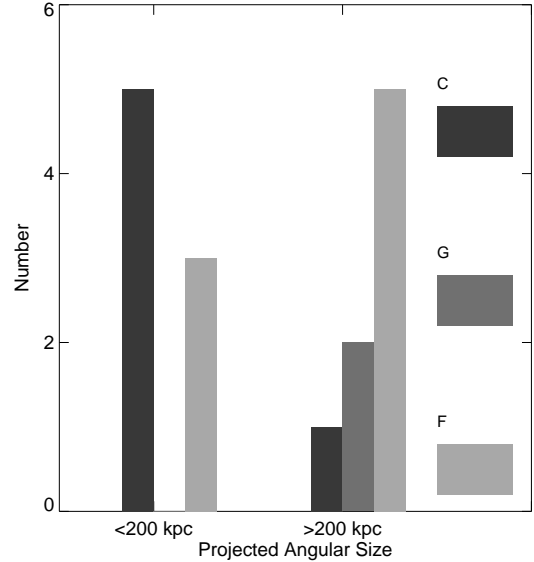,angle=0,width=8cm}

    \caption{Bar plot showing the dichotomy in projected angular size
    for sources in field (F), compact groups (G) and cluster (C)
    environments.}

    \label{fig:bar}

  \end{center}

\end{figure}

Figure~\ref{fig:bar} shows clustering about our RLQs based on the
projected angular size of the source. Of the 21 sources, those whose
environments could not be classified are excluded. MRC~B0346--279 and
MRC~B1055--242 are also removed because they are unresolved
flat-spectrum objects and therefore yield information only on their
cores, the lobes being either too small or too diffuse to be
observed. MRC~M1355--236 is classified as a cluster in this analysis
because it does not satisfy the modified Hickson requirements to
qualify as a compact group. All other objects are classed according to
column 6 of Table~\ref{tab:bgq}.

Figure~\ref{fig:bar} suggests that clusters of galaxies are often
associated with smaller steep-spectrum radio sources. The Binomial
probability of obtaining this distribution at random from the fields
detailed in Figure~\ref{fig:bar} is $10\%$.

The source size/clustering relation cannot be one-to-one. We would
expect to see small sources in poorer environments for two
reasons. Firstly, the age of the source must be a factor. A young
radio source cannot tell us anything about the environment it has yet
to expand into. Secondly, projection effects may serve to foreshorten
large-scale radio structures of sources in poor environments,
weakening the correlation between angular size and clustering. This
second effect will not be large as we exclude core dominated objects
so removing the most extremely foreshortened objects from the sample.

\begin{table}

  \caption{Characteristics of radio sources with short lobe to long
    lobe ratios of$\leq 0.67$ in our sample. The shorter lobe side is
    indicated by the letter in the second column.}

\begin{center}

\begin{tabular}{lllc}

\hline\hline
MRC Object & Approx lobe & Cluster & F/G/C\\
 & size ratio & side & \\
\hline

B0058--229 & 0.67N & N & G \\

B0209--237 & 0.59S & S & C \\

B0450--221 & 0.67NW & NE & C \\

B0941--200 & 0.30S & S & G \\

B1208--277 & 0.55N & S & C \\

B1301--251 & 0.59W & NE & C \\

\hline

\end{tabular}
\end{center}
\label{tab:ass}
\end{table}

There is a correlation between clustering about our radio sources and
their asymmetry. The short-to-long lobe length ratios of all sources
were determined by calculating the distance from the optical
identifications to the peak in the radio in each lobe. We find that
all sources with lobe length ratios $\leq 0.67$ are associated with
clustering (Table~\ref{tab:ass} ). Of sources that display less
asymmetry, only 5 of 15 show signs of clustering. This suggests that a
source's asymmetry is a strong function of the other galaxies in its
environment.

Of our asymmetric sources, the shortest lobe points toward the main
body of clustering galaxies in only half of the cases. If the lobe
length asymmetry is caused by density gradients in an ICM, this would
indicate that the galaxy and gas distributions are not necessarily
closely related. However, we note that of the six sources in
Table~\ref{tab:ass}, the short lobe and cluster coincide for the three
lower redshift sources. Deeper images of sources above $z=0.8$ may
reveal a different galaxy distribution. Foreshortening of bent sources
can also complicate matters. Nevertheless, the high correspondence of
asymmetry and clustering indicates that cluster ICM affects radio
structure considerably.

Cluster searches based on radio-loud AGN selection are currently
highly-efficient means of finding high-redshift systems of
galaxies. Evidence from analysis of radio morphology suggests that
selecting sources based on radio size and asymmetry can improve the
efficiency of these searches still further.

\section{Conclusions}

We have described the environments of a representative sample of RLQs
in terms of the colours of galaxies which surround them. Of the 21
sources, we have identified clustering of galaxies around 8 of
them. Three further sources show signs of clustering but require more
accurate observations to confirm. Two other sources display strong
evidence for a lower-redshift galaxy systems in their fields. The
other 8 quasars in the sample reside in environments indistinguishable
from the field.

As found at lower redshifts, powerful low-frequency-selected
radio-loud quasars at redshift $0.6<z<1.1$ exist in a wide
variety of environments, from field through compact groups to rich
clusters. There is no evidence for evolution in clustering richness as
we progress to higher redshifts.

The colours of galaxies in the RLQ-selected clusters are suggestive of
passively-evolving ellipticals with no more than a small amount of
on-going star formation. The quasar is not always directly centred on
any overdensity, nor is clustering confined to within 0.5 Mpc of the
active galaxy. This contrasts with clusters at lower redshift where
quasar host galaxies, massive ellipticals, appear to reside at the
centre of a relaxed galaxy distribution. It also conflicts with the
assertion presented above that the environments of these sources do
not vary with redshift.

The positions of the quasars and the distribution of the galaxies
suggest that these clusters are not virialised. This indicates that
massive elliptical galaxy formation is favoured in overdense regions
or virialised subclumps and that the old, red galaxies are in place
before the cluster becomes relaxed.

We have shown that using colour information to describe environment is
a more precise way than quantifying the clustering based on
observations made in a single filter. Indeed, it avoids several of the
pitfalls associated with these clustering statistics; namely the
increased cosmic variance at these faint magnitude levels, the offset
of the RLQ from the centre of the galaxy distribution and signals
caused by foreground objects. We predict that where $B_{gq}$ has been
used to quantify clustering above $z=0.6$, it will systematically
underestimate the clustering richness of true clusters. In about
$10\%$ of cases at $z \sim 1$, $B_{gq}$ may be influenced by the
presence of foreground systems.

The smallest extended sources show a strong association with clusters
of galaxies. Asymmetric sources are also more likely to be classified
as being in clustered environments (though not necessarily in the
richer systems). Discrimination on size and asymmetry results in an
extremely efficient way of selecting high-redshift groups and
clusters. Despite the obvious effectiveness of colour selection in
detecting these systems, follow-up 8m imaging and spectroscopy is
still required to determine the parameters of the systems.

\section*{Acknowledgments}

JCB acknowledges support from the Royal Society through a University
Research Fellowship, and also funding which was provided by NASA
through Hubble Fellowship grant \#HF-01103.01-98A from the Space
Telescope Science Institute, which is operated by the Association of
Universities for Research in Astronomy, Inc., under NASA contract
NAS5-26555.

This work was based on observations collected at the European Southern
Observatory, Chile. The ESO data was obtained as part of programmes
56.A-0733, 58.A-0825 and 60.A-0750. This research has made use of the
NASA/IPAC Extragalactic Database (NED) which is operated by the Jet
Propulsion Laboratory, California Institute of Technology, under
contract with the National Aeronautics and Space Administration.

\appendix

\section{Single-filter clustering statistics}
\label{app:bgq}

\subsection{Galaxy-quasar cross-correlation amplitude, $B_{gq}$}

The amplitude of the spatial cross-correlation function has been used
as a measure of galaxy excess by many investigators, initially Seldner
\& Peebles (1978)\nocite{seldner78} and Longair \& Seldner
(1979)\nocite{longair79}. Its application to probe the environments of
AGN at various redshift is also well established
\cite{yee87,ellingson91,wold00,mclure01}. A detailed derivation can be
found in Longair \& Seldner (1979)\nocite{longair79}. 

The spatial cross-correlation function, $\xi(r)$ measures the excess
probability of finding a companion galaxy at a distance $r$ from the
quasar. It is found to be well fit by the form,

\[
\xi(r) = B_{gq} r^{-\gamma}
\vspace{-0.5mm}
\]

\noindent
where $B_{gq}$ is the spatial cross-correlation amplitude and $\gamma$
is the power-law index. The value of $\gamma$ is usually assumed to be
1.77, the value for the low-redshift galaxy-galaxy correlation
function as found by Groth \& Peebles (1977)\nocite{groth77}.  In
order to determine $B_{gq}$ from our data, we must first find the
angular cross-correlation amplitude, $A_{gq}$, from the angular
counterpart to the spatial cross-correlation function,

\[
\omega(\theta)=A_{gq}\theta^{1-\gamma}
\]

\noindent
For small $\theta$ the amplitude of the angular cross-correlation
amplitude can be estimated from the data using the relation:

\[
A_{gq}=\frac{N_{tot}-N_{b}}{N_{b}} \frac{{3-\gamma}}{2} \theta^{\gamma-1}
\]

\noindent
where $N_{tot}$ is the number of galaxies within a circle of radius
$\theta$ centred on the quasar corresponding to 0.5 Mpc at the
redshift of the quasar. $N_{b}$ is the expected number of background
galaxies within $\theta$. The spatial cross-correlation amplitude is
then obtained from $A_{gq}$ using:

\begin{equation}
B_{gq}=\frac{N_{g} A_{gq}}{\Phi(m_{lim},z)I_{\gamma}} d_{A}^{\gamma-3}
\label{eqn:bgq}
\end{equation}

\noindent
where $N_{g}$ is the average surface density of background galaxies
and $d_{A}$ is the angular diameter distance to the
quasar. $I_{\gamma}$ is an integration constant which has the value
3.78 for $\gamma$ = 1.77, and $\Phi$($m_{lim},z$) is the luminosity
function integrated down to the completeness limit at the quasar
redshift. This gives an estimate of the number of galaxies per unit
volume to the limiting magnitude and introduces a degree of redshift
independence into the statistic.

All of our images have a large enough field of view so that we can use
the area further than 0.5 Mpc from the quasar to determine a ``field
density''. We then used this to predict a value of $N_{b}$, the number
of objects expected within 0.5 Mpc of the quasar.

For the Schechter parameters, we follow the prescription given in W00
and use $\alpha=0.89$, $\phi_{V}^{*}=0.0065$, $\phi_{R}^{*}=0.0072$,
$\phi_{I}^{*}=0.0052$ and $M_{V}^{*}=-21.04$, $M_{R}^{*}=-22.08$,
$M_{I}^{*}=-22.65$. In order to probe as far down the luminosity
function as possible, the deepest image in any particular field was
used to calculate $B_{gq}$. This means where we had the choice, we
first used the EFOSC $I$ band, then the EFOSC $R$ and then the CTIO
$I$. In the case of the Taurus data which does not overlap with the
EFOSC and CTIO data we had no choice, and the $I$ band was used to
calculate $B_{gq}$. The errors were calculated using the conservative
prescription of Yee \& Lopez-Cruz (1999)\nocite{yee99}:

\[
\frac{\Delta B_{gq}}{B_{gq}} = \frac{[(N_{tot}-N_{b})+1.3^2 N_{b}]^{1/2}}{N_{tot}-N_{b}}
\]

\noindent
where the variables have the same meaning as before.

\subsection{The Hill \& Lilly Quantity, $N_{0.5}$}

The quantity $N_{0.5}$ \cite{hill91} provides a measure of spatial
clustering which, unlike $B_{gq}$, does not depend on an estimate of
the luminosity function at the redshift of the quasar. For a galaxy
with apparent magnitude $m_{g}$, $N_{0.5}$ is defined as the number of
excess galaxies within a projected 0.5 Mpc with magnitudes in the
range $m_{g}$ to $m_{g} + 3$. When looking at quasars, light from the
AGN will dominate light from the host galaxy, making a direct
measurement of $m_{g}$ impossible. $m_{g}$ is therefore estimated from
the magnitude-redshift relation for radio galaxies of Eales
(1985)\nocite{eales85}:

\begin{equation}
m_{g}(R)=21.05+5.3 \mathrm{log} \, z
\label{eqn:bcg}
\end{equation}

\noindent
valid for $19.0 < m_{g} < 21.3$. Magnitudes for objects in the quasar
fields which we imaged in colours other than $R$ (13 fields), were
adjusted to $R$ using the colour relationships for redshifted E/S0
galaxies from Fukugita \etal (1995)\nocite{fukugita95}. The number
density of expected objects was calculated from the region outside 0.5
Mpc as per the $B_{gq}$ formulation and the error from the $\sqrt{N}$
of this number. In several cases (indicated on Table \ref{tab:bgq}),
the images are not deep enough for a complete catalogue of objects to
$m_{g} + 3$. In these cases $N_{0.5}$ is calculated to the
completeness limit of the data as given Table~\ref{tab:obs}. For a
fuller description of $N_{0.5}$, again see W00.

\begin{figure}

  \begin{center}

    \epsfig{file=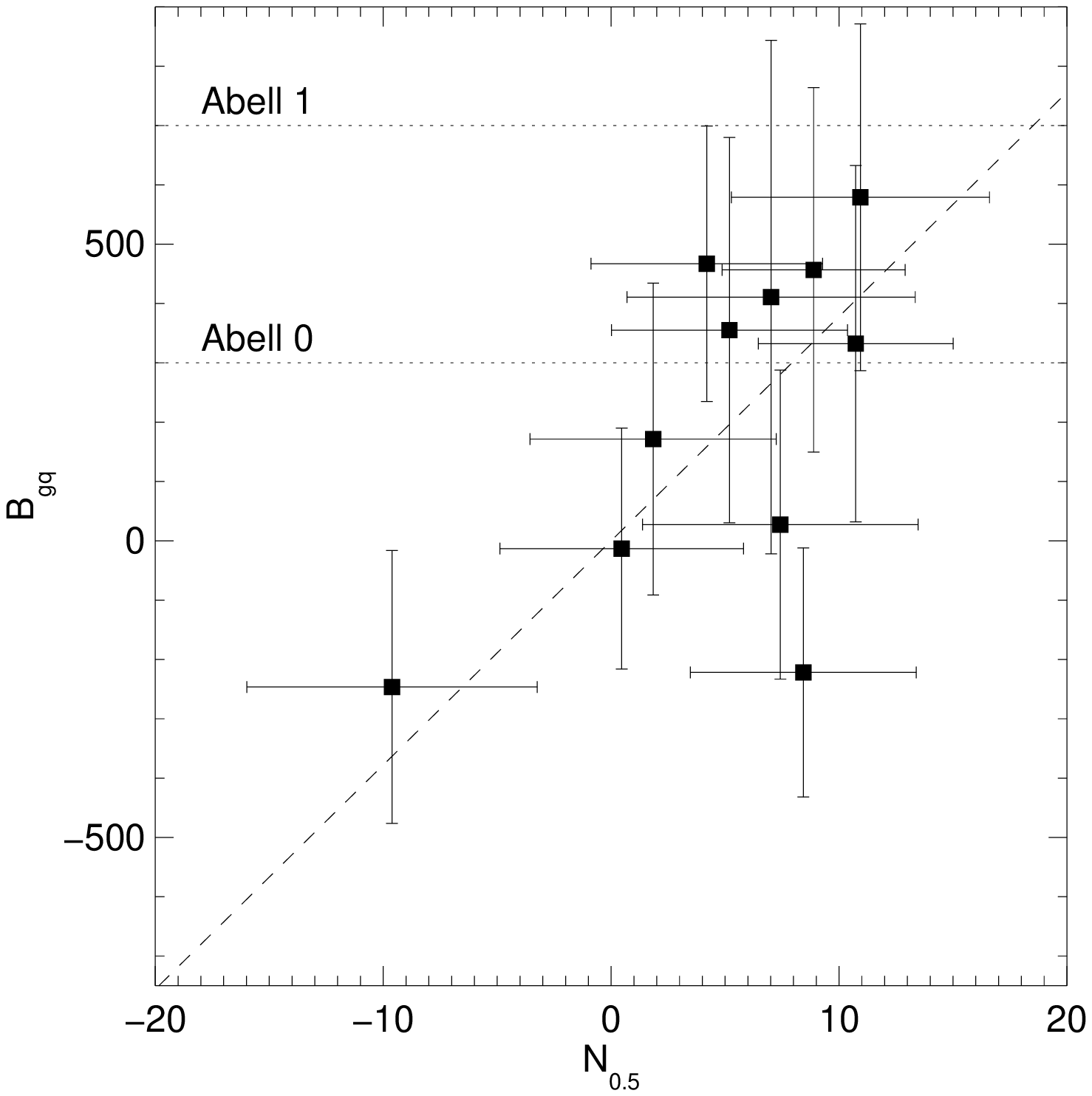,angle=0,width=8cm}

    \caption{Graph of $B_{gq}$ {\em vs.} $N_{0.5}$ for the 11 fields
    in our sample which are complete to $m_{g} + 3$ for the $N_{0.5}$
    calculation. The dashed line is the $N_{0.5}$ - $B_{gq}$ relation
    of W00, $B_{gq} = 37.8 N_{0.5}$. The horizontal lines indicate the
    normalisation between $B_{gq}$ and Abell classes as quoted in
    McLure \& Dunlop (2001).}

   \label{fig:bgqvsn05}

  \end{center}

\end{figure}

Figure~\ref{fig:bgqvsn05} shows the relation between the spatial
cross-correlation amplitude and the Hill and Lilly quantity for the
fields that we have values of $N_{0.5}$ calculated using the full
$m_{g}$ to $m_{g} + 3$ magnitude range. We see a close relationship
between the quantities as they both probe the clustering of the same
faint objects in the fields of the quasars. As Figure~\ref{fig:bgqvsn05}
shows, there is agreement which is consistent with the relationship of
$B_{gq} = 37.8 N_{0.5}$ quoted in W00.

\bibliography{/export/home/jbarr/latex/refs}
\label{lastpage}

\end{document}